\documentclass{aa}  

\usepackage{graphicx}
\usepackage{txfonts}
\usepackage{natbib}
\usepackage{multirow}
\usepackage{txfonts}
\usepackage{tabularray}
\usepackage{subcaption}
\usepackage{footnotehyper}
\usepackage{array}
\usepackage{booktabs}
\usepackage{caption}
\usepackage{subcaption}
\usepackage{gensymb}
\usepackage{amssymb}
\usepackage{booktabs}
\usepackage{url}
\usepackage{multirow}
\usepackage{siunitx}
\usepackage{wrapfig}
\usepackage{lscape}
\usepackage{rotating}
\usepackage{longtable}
\usepackage{epstopdf}
\usepackage{caption}
\usepackage{txfonts}
\usepackage{natbib}
\usepackage{subcaption}
\usepackage{longtable}
\usepackage{array}
\usepackage{caption}
\usepackage{tablefootnote}
\newcolumntype{y}[1]{>{\centering\let\newline\\\arraybackslash\hspace{0pt}}p{#1}}
\bibpunct{(}{)}{;}{a}{}{,} 
\begin{document} 
\setlength{\abovedisplayskip}{20pt}
\setlength{\belowdisplayskip}{20pt}
\setlength{\textfloatsep}{0.2cm}

   \title{A bipolar structure and shocks surrounding\\ the stellar-merger remnant V1309 Scorpii}


   \author{T. Steinmetz\inst{1}\fnmsep\thanks{email: thomas@ncac.torun.pl}
          \and
          T. Kami\'{n}ski\inst{1}
          \and
          M. Schmidt\inst{1}
          \and
          A. Kiljan\inst{2}
          }

   \institute{Nicolaus Copernicus Astronomical Center, ul. Rabia\'{n}ska 8, 87-100 Toru\'{n}, Poland
   \and
   Faculty of Physics, University of Warsaw, ul. Pasteura 5, 02-093 Warsaw, Poland}

   \date{Received XXX, accepted YYY}

  \abstract
   {V1309 Sco is an example of a red nova, a product of the merger between non-compact stars. V1309 Sco is particularly important within the class of red novae due to the abundance of photometric data of the progenitor binary before the merger.}
   {We aim to investigate the spatio-kinematic and chemical properties of the circumstellar environment, including deriving the physical conditions  and establishing the origins of the different circumstellar components.}
   {We use radiative transfer modelling of molecular emission in sub-mm spectra to examine the properties of the molecular gas, and use forbidden line diagnostics from optical spectra to constrain electron density and temperature using forbidden line diagnostics. We compare line intensities from shock models to observations to look for and constrain shocks.}
   {We derive a new kinematical distance of 5.6 kpc to the source. The detection of ro-vibrational H$_2$ and sub-mm HCO$^+$ emission in 2016 and 2019, respectively, indicate active shock interactions within the circumstellar environment. The velocity profiles of both H$_2$ and HCO$^+$, as well as the moment-1 maps of sub-mm CO and $^{29}$SiO, indicate a bipolar structure that may be asymmetric. The sub--mm and optical molecular emission exhibits temperatures of 35--113 and 200 K, respectively, whilst the atomic gas is much hotter, with temperatures of 5--15 kK, which may be due to shock heating.}
   {The detection of a bipolar structure in V1309 Sco indicates further similarities with the structure of another Galactic red nova, V4332 Sgr. It provides evidence that bipolar structures may be common in red novae. All collected data are consistent with V1309 Sco being a kinematically and chemically complex system.}
   

   \keywords{Astrochemistry -- Stars: Individual: V1309 Scorpii -- Stars: winds, outflows -- Stars: Imaging -- Stars: mass-loss -- circumstellar matter}

   \maketitle


\section{Introduction}
Red novae are a class of stellar eruptions believed to be caused by the merger of two non-compact stars \citep{soker2003,soker2006,tylenda2006,soker2007}. Red novae are characterised by low temperatures ($\sim$2000 K), multi-peaked light curves with an intermediate peak luminosity between those typical of classical and supernovae, and high molecular abundances and dust. Sources such as V838 Mon \citep{bond2003, kaminski2015}, V4332 Sgr \citep{kaminski2010v4332, kaminski2011v4332, kaminski2013v4332, tylenda2015v4332}, BLG-360 \citep{tylenda2013blg360} and CK Vul \citep{kaminski2020ckvul, kaminski2021ckvul} have all been previously identified as red novae, the latter of which is the oldest known red nova and had previously been misidentified as a classical nova. Multiple red novae have also been discovered in other galaxies \citep[][and references therein]{pastorello2019}.

V1309 Sco erupted in 2008 \citep{nakano2008} and remains the only example of a Galactic red nova whose progenitor has been regularly observed. The OGLE survey \citep{udalski2003} photometrically observed the progenitor binary system over multiple years, revealing that the binary system was made up of two K-type sub-giants with masses of (0.5--0.9) and (1.1--1.3) M$_{\odot}$. Its orbital period of P$\approx$1.4 days was exponentially decreasing when approaching the common envelope (CE) phase and the merger event \citep{stepien2011,tylenda2011, pejcha2014}. This is the only example to date of a binary system of non-compact stars observed during the spiral-in phase of a merger event. V1309 Sco therefore provides a useful testing ground for establishing connections between the properties of the progenitor binary and the merger remnant.

The nature of the stellar remnant is currently unknown due to steady formation of dust obscuring the star, starting from two years after the merger event \citep{nicholls2013,mccollum2013}, but the origins and nature of the circumstellar environment (CSE) surrounding the remnant are of interest. In particular, we wish to understand the evolution of post-merger remnants and find connections between the kinematic structure and chemical markers associated with such events. The CSE has been predicted to come from a variety of sources: L$_2$ mass loss before the merger, circumbinary disk formation, common envelope (CE) phase, merger ejecta and stellar winds \citep[e.g.][]{zhu2013,nandez2014, pejcha2014,pejcha2017, matsumoto2022, macleod2020, macleod2022}.

Several observational studies attempted to infer the structure of the remnant. \citet{kaminski2018} (hereafter \textbf{K18}) revealed a two-component structure in the sub-millimeter/millimeter spectrum of V1309 Sco obtained with the Atacama Large sub-Millimeter Array (ALMA), indicating a velocity gradient in the moment-1 maps of SiO and showing a similar structure to another Galactic red nova, V4332 Sgr, which was shown to exhibit bipolar outflows. \citet{mason2022} independently detect two absorption components in spectra taken soon after the 2008 eruption, which they associate with material ejected both before and during the coalescence. One of these components switches from absorption and emission 25--35 d after the eruption, showing that the circumstellar material evolved rapidly at this time.

The chemical composition is also worthy of study. V1309 Sco is known to be oxygen-rich, and so is abundant in oxygen-bearing molecules. \citet{kaminski2015} reported the detection of CrO in V1309 Sco for the first time, a molecule only detected in V4332 Sgr and in no other classes of objects. The presence of such a rare molecule in red nova remnants indicates that non-standard chemistry occurs in such objects.

This study aims to gain a better understanding of the kinematic and spectroscopic structure of the remnant of V1309 Sco after 2012. In Sects. \ref{section: ALMA} and \ref{section: xshooter}, we present observations, results and analysis of ALMA and XSHOOTER observations, respectively. Section \ref{section: discussion} presents our discussions of the results and Sect. \ref{section: conclusions} presents our conclusions. Also provided are several appendices showing tables of detected lines and bands in both ALMA and XSHOOTER, as well as details on the modelling of optical molecular bands and statistics of H$_2$ shock  models used to examine the shock properties in the CSE.

\section{ALMA}\label{section: ALMA}
We start with presenting the most recent ALMA observations of V1309 Sco.
\subsection{Observations}\label{section: ALMA obs}
V1309 Sco was observed with ALMA in band 7 on 17$^{th}$ January, 26$^{th}$ March and 9$^{th}$ April 2016. The details of these observations are described in detail by \textbf{K18}. Band 7 observations were repeated on 26$^{th}$--28$^{th}$ August 2019 (PI: T. Kami\'{n}ski). The setup used in 2019 had longer baselines of up to 3.6 km and science integration time of $\sim$2.5 hours, compared to maximum baselines of 460 m and a science integration time of $\sim$48 min in 2016. The longer baselines meant a spatial resolution of almost one order of magnitude better than in 2016. A summary of the comparison between the observational epochs is shown in Table \ref{table: ALMA obs}. The 2019 baselines gave a clean primary beam size (FWHM) of 80$\times$60 mas at uniform weighting, whilst the spectral coverage was 342.5--358.1 GHz with a gap between $\sim$346--355 GHz due to the heterodyne setup. The spectral windows were shifted from the setup used for the 2016 observations by $\sim$2 GHz (see Table \ref{table: ALMA obs}) in order to cover the H$^{13}$CN (4--3) line ($\nu_{\textrm{rest}}$=345.34 GHz), which was subsequently not detected. The ALMA data was reduced with CASA pipeline version 4.5.3 using the default calibration script. The source was too faint for self-calibration. The data was imaged using the CASA routine \texttt{tclean}.

As noted in \textbf{K18}, the ALMA field of view covers an unidentified sub-mm source at RA=17$^h$57$^m$32\fs6768, Dec=$-30$\degr43\arcmin14\farcs157 (J2000). It is located $\sim$5\arcsec\ south-west from V1309 Sco and is bright in the 2019 observations. This source is likely an uncategorised background galaxy.

\subsection{Line identification}\label{section: ALMA line ident}
To identify the sub-mm/mm spectral features, we used the CASSIS spectroscopic analysis tool\footnote{http://cassis.irap.omp.eu} \citep{vastel2015cassis} to extract spectroscopic information from the Cologne Database for Molecular Spectroscopy \citep[CDMS;][]{CDMSref} and the Jet Propulsion Laboratory database \citep[JPL;][]{JPLref}. All identified lines are listed in Table \ref{table: ALMA lines}. As the ALMA spectrum is mostly unchanged in 2019 compared to 2016 (see Fig. \ref{fig: ALMA spectrum labelled}), the majority of lines seen in the 2019 spectrum could be identified without major effort. However, due to the aforementioned shift in the spectral windows for the 2019 observations relative to 2016, the SO$_2$ 20(0,20)--19(1,19), 25(3,23)--25(2,24), SO 9(8)--8(7), and $^{28}$SiO 8--7 transitions are not observed in 2019. Instead, $^{29}$SiO 8--7 and SO$_2$ 12(4,8)--12(3,9) are the newly observed lines, not covered in 2016 (Table \ref{table: ALMA lines}). From the moment-0 maps of CO (3--2), the flux has increased from 181.5 Jy/(beam km s$^{-1}$) in 2016 to 226.0 Jy/(beam km s$^{-1}$) in 2019 (26\% increase).

Perhaps the most significant is the identification of HCO$^+$ for the first time in V1309 Sco (Table \ref{table: ALMA lines}). As seen in Fig. \ref{fig: ALMA spectrum labelled}, the HCO$^+$ (4--3) line is blended with the SO$_2$ 10(4,6)--10(3,7) line in 2016 and could not be identified. In 2019 the strength of HCO$^+$ relative to SO$_2$ increased. This observation, coupled with the better spectral resolution of the 2019 ALMA spectrum, allowed HCO$^+$ to be (partially) resolved from SO$_2$.
 
The ALMA spectrum is dominated by SO$_2$ emission, with 20 individual lines identified. The average LSR peak velocity across all identified lines in Table \ref{table: ALMA lines} is --81 km s$^{-1}$.

\begin{table}
\caption{Summary of ALMA observations in 2016 and 2019.}
    \centering
    \renewcommand{\arraystretch}{1.1} 
    \begin{tabular}{lcc}\hline
    Property & Epoch 1 & Epoch 2\\\hline
    Frequency range & 344.2--360.0 & 342.5--358.1 \\
    (GHz)&&\\
    & 17-01-2016 & 26-08-2019 \\
    Dates observed & 26-03-2016 & 27-08-2019 \\
    & 09-04-2016 & 28-08-2019 \\
    &&\\
    \multirow{2}{10em}{Angular resolution (arcsec)} & 0.534 & 0.068 \\&&\\
    \multirow{2}{10em}{Spectral resolution (MHz/bin)} & 7.813 & 1.950 \\&&\\
    \multirow{2}{10em}{Spectral resolution \\(km s$^{-1}$/bin)} & 6.54 & 1.63 \\&&\\
    \multirow{2}{10em}{Science integration time (s)} & 2856 & 9018 \\&&\\
    Baselines (m) & 15--460 & 38--3637 \\
    E$_{\rm up}$ (K) & 48.48--521.00 & 48.48--581.92\\\hline
    \end{tabular}
\tablefoot{E$_{\rm up}$ is only for SO$_2$ transitions covered by the frequency range.}
\label{table: ALMA obs}
\end{table}
\subsection{Source size and distance}\label{section: ALMA size+distance}
Using moment-0 maps combined from the full measurement sets, we used the CASA routine \texttt{imfit} to fit elliptical Gaussians to the spatial distribution of the source in each epoch. The images used represent all emission within all spectral windows. The resulting source sizes (deconvolved from the main beam) measured were 178 ($\pm$21) $\times$ 125 ($\pm$36) and 87.9 ($\pm$1.8) $\times$ 59.8 ($\pm$1.7) mas respectively for epochs 1 and 2. As a sanity check, the images were cleaned using natural and uniform weightings to compare the results, with the different weightings providing the same results per epoch within 1$\sigma$ errors. As each epoch had different beam sizes, we smoothed the epoch 2 data to the same beam size (0\farcs51$\times$0\farcs45) as epoch 1. The \texttt{imfit} results provided the same results as those previously found for the unsmoothed epoch 2 data. At face value, comparing the source size in each epoch would indicate that the source has contracted between 2016 and 2019. However, we find this highly unlikely and rather consider the uncertainties in epoch 1 measurements to be largely underestimated.

Using the CO (3--2) emission from the 2019 observations, we estimate the kinematical distance of V1309 Sco. The CO (3--2) distribution has a semi-major axis of 45.5 mas, assumed to be the semi-major FWHM/2 of the 2D Gaussian fitted to the CO (3--2) spatial distribution using \texttt{imfit}. We assume that the radial velocity is equal to the velocities observed in the far wings of the CO (3--2) line, and so take half the baseline width as the tangential velocity in one direction, which turned out to be $\varv_e$=232.5 km s$^{-1}$. The light curve peak of V1309\,Sco occurred on 2.5 September 2008 \citep[julian date 2 454 712;][]{tylenda2011} and the first date of the epoch 2 observations was 26 August 2019. We assume that the CO gas was ejected at the point of the merger, and so $t$=4012 d (=3.464$\times$10$^8$ s) had elapsed. This gives a physical radius of the CO emission region. This also provides an upper limit on the distance, as the cold CO gas may have been ejected prior to the merger, meaning the time elapsed is potentially much longer and the CO emission radius is therefore larger. Converting the FWHM/2=0\farcs0455 to a value\footnote{Here, $\sigma$ refers to the standard deviation of the Gaussian profile fitted to the source, calculated using $\sigma$=FWHM/2$\sqrt{2\ln(2)}$} of 5$\sigma$ (=0\farcs097) and adopting this as the source size, $\theta$, we derive the distance $d=(\varv_e \times \textrm{t})/\tan(\theta)$. The kinematic distance we determine is therefore 5.6 kpc, which is consistent with the distance estimate presented in \citet{mason2022}. This corresponds to a distance of the most extended CO gas from the central star of $\sim$540 AU.

\subsection{Kinematics and structure}\label{section: ALMA kinematics}
\subsubsection{Moment maps}\label{section: moment maps}
As the dominant emission originates from the CO (3--2) and $^{29}$SiO (8--7) lines, both were used to probe the spatio-kinematic structure in 2019. Moment-0 and 1 maps were constructed using the CASA routine \texttt{immoment}. We extracted the emission across the same velocity range of --310 to 160 km s$^{-1}$. Any pixels with values < 5$\sigma$ than the root--mean--square (rms) noise were neglected. The CO moment-1 map from 2019 is shown in Fig. \ref{fig: moment maps} (top panel). 
\begin{figure*}
    \centering
    \includegraphics[scale=0.12]{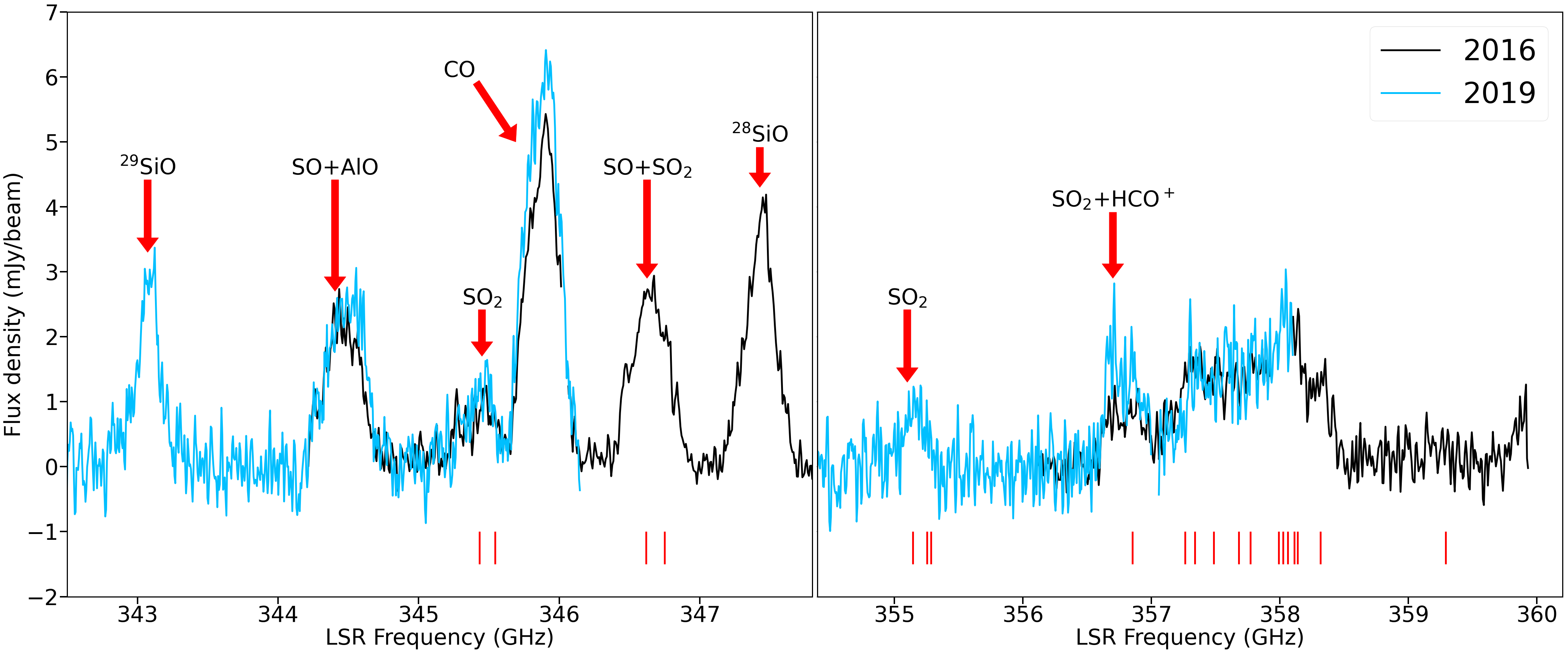}
    \caption{ALMA band 7 spectra of V1309 Sco. The reference frame is the local standard of rest (LSR). Black shows the 2016 spectrum, and blue shows the 2019 spectrum. The key lines are indicated by the labels. Red vertical lines indicate the position of the multiple identified SO$_2$ lines shifted to the observed positions. The 2019 spectrum was extracted from data smoothed to the same beam size as the 2016 data. Both spectra represent the entire source. Both spectra are smoothed.}
    \label{fig: ALMA spectrum labelled}
\end{figure*}
\begin{figure}[h!]
\centering
\includegraphics[width=0.5\textwidth]{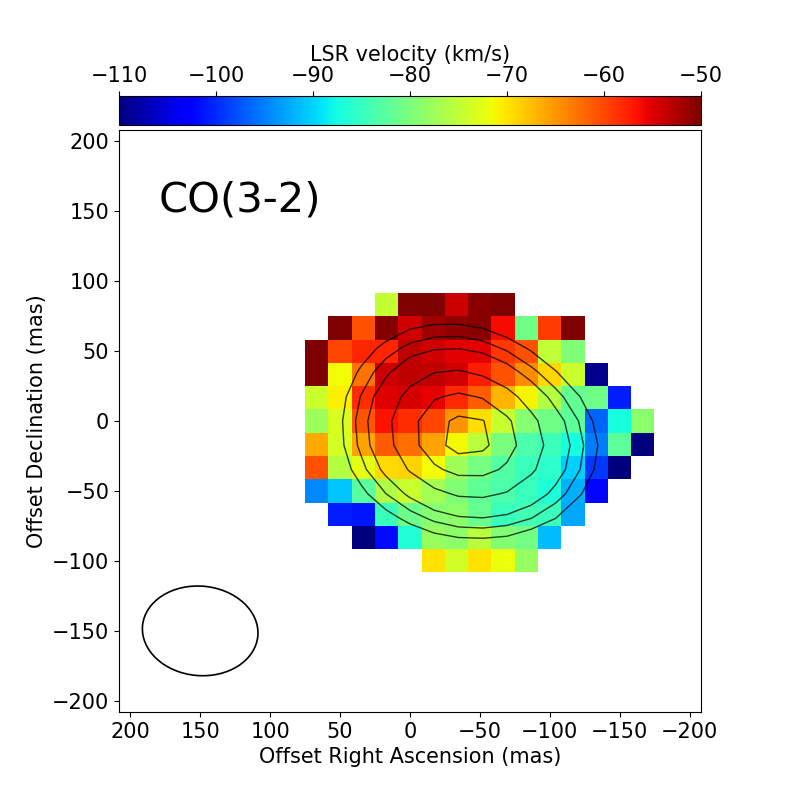}
\includegraphics[width=0.5\textwidth]{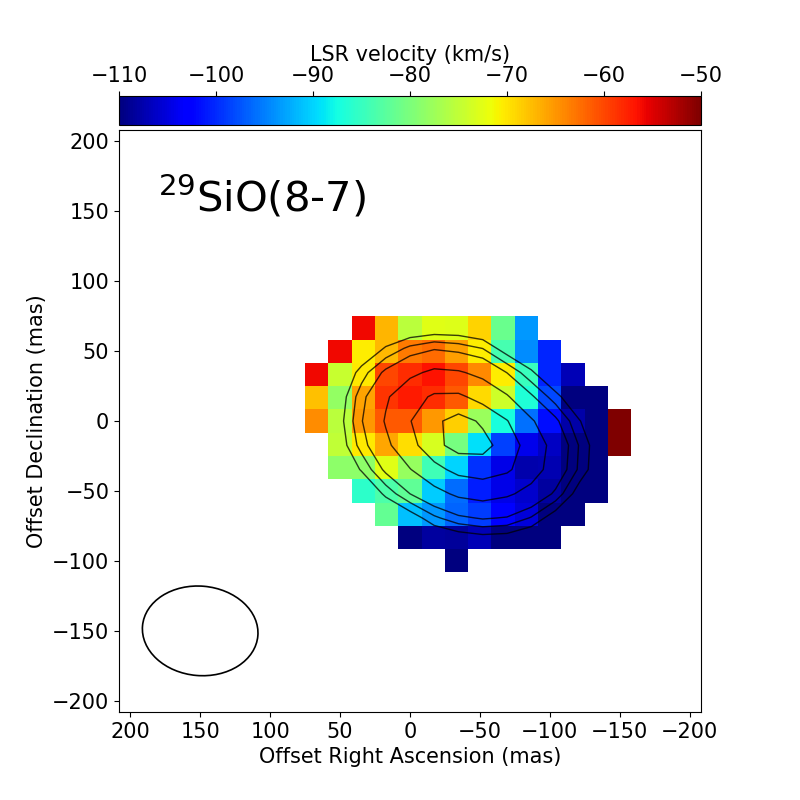}
    \label{subfig: 2019 mom 1}
    \caption{\textbf{Top:} Moment-1 map of the 2019 CO (3--2) line, extracted across --310 to 160 km s$^{-1}$. Black contours represent 5, 10, 20, 40, 80 and 95\% of the peak CO (3--2) flux. \textbf{Bottom:} Moment-1 map of the $^{29}$SiO (8--7) line, extracted across --260 to 160 km s$^{-1}$. Black contours represent 5, 10, 20, 40, 80 and 95\% of the peak $^{29}$SiO (8--7) flux. The colourbars cover the same velocity range for both maps. Pixels with values below 5$\sigma$ noise level were blanked in all maps.}
    \label{fig: moment maps}
\end{figure}
Even though the overall source structure is still largely unresolved in 2019, the better spatial resolution reveals a velocity gradient spanning from north-east to south-west, with the north-east region moving away from the observer. The gradient axis approximately matches the direction of a slight elongation seen in the CO (3--2) moment-0 maps.

We also constructed moment-0 and 1 maps of the $^{29}$SiO (8--7) line (Fig. \ref{fig: moment maps}, bottom panel). The moment-0 map also shows a nearly point-like structure, as seen in CO (3--2). The $^{29}$SiO (8--7) moment-1 map shows an identical structure to the observed velocity gradient in the $^{28}$SiO (8--7) line from 2016 \citep[][their Fig. 9]{kaminski2018}. At Solar composition, $^{29}$Si constitutes only 5\% of the main isotope $^{28}$Si, and thus $^{29}$SiO emission should be optically thinner than $^{28}$SiO (cf. Sect. \ref{section: ALMA tau}). The velocity structure of the $^{29}$SiO (8--7) emission shows a redshifted lobe in the north-east, with less extreme velocities at extended distances. This is inconsistent with a rotating sphere, which would show the most extreme velocities in the moment map at the most extended distances.


\subsubsection{Radiative transfer modelling}\label{section: RT modelling}
To characterise the molecular gas observed with ALMA, we used the local thermodynamic equilibrium (LTE) radiative transfer tool available in CASSIS. We modelled all detected molecules (shown in Table \ref{table: ALMA lines}) as well as $^{34}$SO$_2$.

\begin{figure*}
    \centering
    \includegraphics[width=\textwidth]{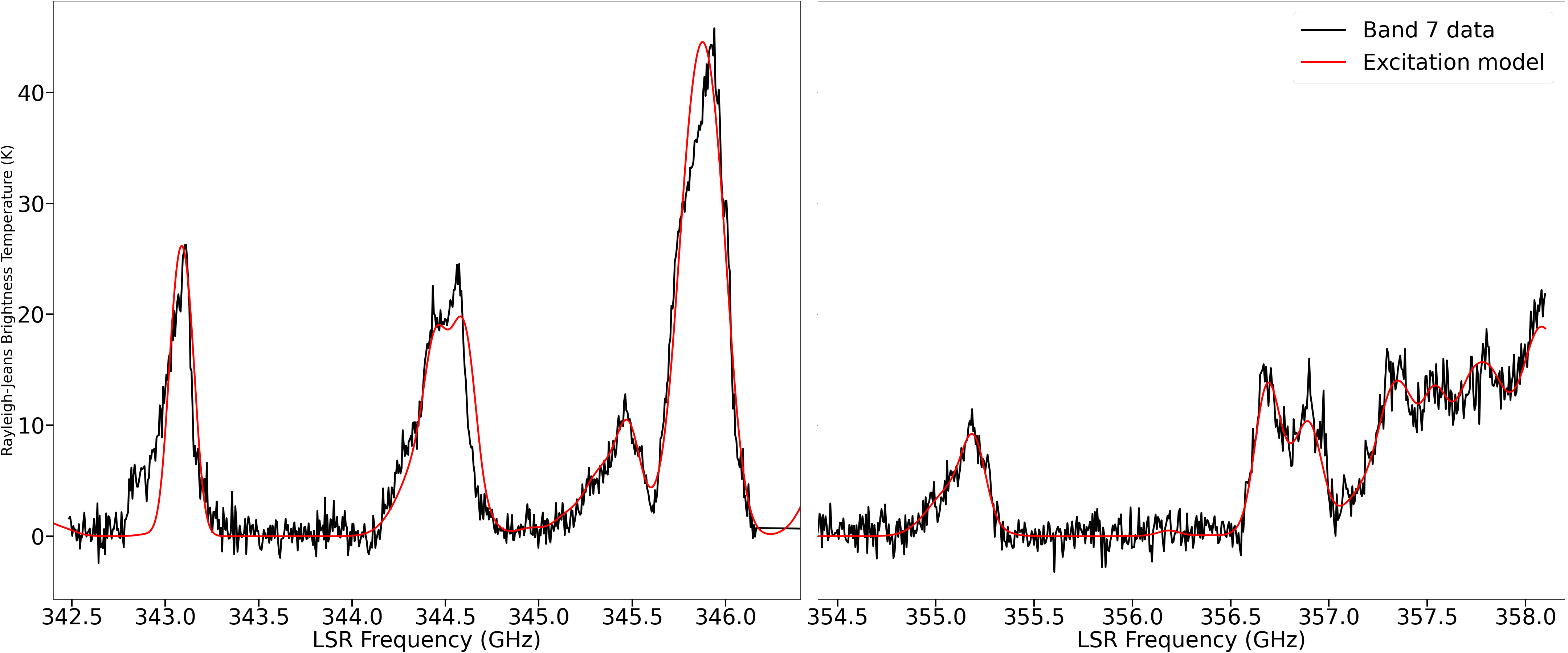}
    \caption{ALMA band 7 spectrum (black) with the best fitting LTE model overlaid (red).}
    \label{fig: LTE model}
\end{figure*}

\begin{table}
    \caption{Parameters of SO$_2$ gas for both ALMA epochs.} \label{table: chi^2}
    \renewcommand{\arraystretch}{1.1} 
    \begin{tabular}{c|cc|cc}\hline
& \multicolumn{2}{c|}{2016} & \multicolumn{2}{c}{2019} \\
& BC & NC & BC & NC \\\hline
$N$ (cm$^{-2}$) & 9.51$\times$10$^{16}$ & 2.56$\times$10$^{16}$ & 4.55$\times$10$^{17}$ & 1.63$\times$10$^{17}$ \\
$T_{\rm ex}$ (K) & 113 $\pm$ 9 & 113 $\pm$ 11 & 35 $\pm$ 10 & 97 $\pm$ 3 \\
$\varv_{\rm LSR}$ (km s$^{-1}$) & --95 & 53 & --43 & --123 \\
FWHM & \multirow{2}{*}{160} & \multirow{2}{*}{90} & \multirow{2}{*}{233} & \multirow{2}{*}{112} \\
(km s$^{-1}$) & & &  \\\hline
    \end{tabular}
    \tablefoot{BC and NC refers to the broad and narrow component, respectively. For the 2016 observations, the FWHM and $\varv_{\rm LSR}$ were both fixed to values derived in \textbf{K18} for both components.}  
\end{table}
The initial model fitting was performed only on SO$_2$ using a Monte-Carlo Multi-Chain (MCMC) $\chi^2$ fitting algorithm, simulating 20 detected SO$_2$ lines to get the best estimates of the SO$_2$ kinetic temperature and column density. The upper energy level range of the SO$_2$ transitions cover 48--521 K, slightly wider than the range covered by the 2016 observations. This should result in better constraints of the excitation temperature in 2019. We then assume that emission of all detected molecules is produced in the same LTE conditions, meaning we assume the same temperature for all molecules for each gas component. We model two gas kinematic components for all molecules except for $^{29}$SiO, HCO$^+$, and CO, which are better represented by a single Gaussian. The key change made to fit the model was the gas column density for each molecule. The results for the SO$_2$ fitting and the overall radiative transfer modelling are presented in Tables \ref{table: chi^2} and \ref{table: LTE model}, respectively, and the best model is shown in Fig. \ref{fig: LTE model}.

\begin{table}
    \caption{Column densities (in cm$^{-2}$) of best fitting model for each species constrained by the LTE modelling in CASSIS.}
    \renewcommand{\arraystretch}{1.1} 
    \begin{tabular}{c|cc|cc}\hline
    Species & \multicolumn{2}{c|}{Epoch 1} & \multicolumn{2}{c}{Epoch 2}\\
    & BC & NC & BC & NC\\\hline
    SO$_2$ & 9.51$\times$10$^{16}$ & 2.56$\times$10$^{16}$ & 4.55$\times$10$^{17}$ & 1.63$\times$10$^{17}$\\
    $^{34}$SO$_2$ & (4.25$\times$10$^{15}$) & (1.15$\times$10$^{15}$) & 2.04$\times$10$^{16}$ & 7.85$\times$10$^{15}$\\
    SO & 1.5$\times$10$^{15}$ & 1.6$\times$10$^{15}$ & 7.00$\times$10$^{16}$ & 5.66$\times$10$^{16}$\\
   $^{28}$SiO & 3.00$\times$10$^{15}$ & - & (6.12$\times$10$^{14}$) & -\\
    $^{29}$SiO & (1.53$\times$10$^{14}$) & - & 1.20$\times$10$^{16}$ & -\\
    CO & 6.00$\times$10$^{18}$ & - & 1.20$\times$10$^{19}$ & -\\
    AlO & 1.25$\times$10$^{15}$ & 1.00$\times$10$^{15}$ & 4.80$\times$10$^{15}$ & 3.40$\times$10$^{15}$\\
    AlOH & 1.00$\times$10$^{15}$ & - & 1.00$\times$10$^{15}$ & -\\ 
    HCO$^+$ &2.60$\times$10$^{14}$ & - & 1.45$\times$10$^{15}$ & -\\\hline
    \end{tabular}
    \tablefoot{$^a$CO was modelled with $\varv_{\textrm{LSR}}$ equal to --72.5 km s$^{-1}$ and a FWHM of 200 km s$^{-1}$, as in \textbf{K18}. Likewise, $^{28}$SiO was modelled with $\varv_{\textrm{LSR}}$ equal to --82 km s$^{-1}$. Epoch 1 data was from the best fitting model measured in the re-analysis of the data, and not taken from the literature. Values in brackets have been calculated from natural (Solar) isotopic abundances.}
    \label{table: LTE model}
\end{table}
We therefore abbreviate these two components to the broad component (\textbf{BC}) and narrow component ({\bf NC}). BC is found at the less negative velocity of --43 km s$^{-1}$ and has a larger line width of 233 km s$^{-1}$. NC is at a velocity of --123 km s$^{-1}$ and has a FWHM of 112 km s$^{-1}$.

The detection of HCO$^+$ is important for recognizing the late role of outflows in the CSE. However, the HCO$^+$ (4--3) line is heavily blended with SO$_2$. We use the model of SO$_2$, shown in Table \ref{table: LTE model} and Fig. \ref{fig: LTE model} to recover the intrinsic line profile of HCO$^+$ by subtracting the SO$_2$ model from the ALMA 2019 spectrum. The profile is further discussed in Sect. \ref{discussion: outflows}.

\subsubsection{Optical depth}\label{section: ALMA tau}
The CASSIS software applies a limited correction on line saturation effects. It is important to know the opacity of the strongest emission features, since we use these to trace the source structure. Therefore, the optical depth ($\tau$) for CO, $^{28}$SiO, and $^{29}$SiO was calculated as
\begin{equation}
    \tau = \frac{c^3}{8\pi\nu^3} \frac{1}{\Delta v} \frac{N}{Z} A_{ul}g_ue^{\frac{-E_{\rm up}}{T}}\left[e^{\frac{h\nu}{kT}}-1\right],
    \label{eq: tau}
\end{equation}
where $N$ is the column density, $T$ is the gas temperature, $\Delta v$ is the FWHM of the modelled molecular lines, $\nu$ is the rest wavelength and all other constants are associated quantum values of each transition \citep{goldsmith1999}. All measurable values ($N$, $T$, $\Delta v$) were obtained from the LTE radiative transfer modelling described in Sect. \ref{section: RT modelling}. The results are presented in Table \ref{table: tau}.
\begin{table}
    \caption{Calculated optical depths of CO, $^{28}$SiO, and $^{29}$SiO.}
    \centering
    \renewcommand{\arraystretch}{1.1} 
    \begin{tabular}{cccc}\hline
    Property & CO &$^{28}$SiO & $^{29}$SiO \\\hline
    Frequency (GHz)& 345.796 & 347.330 & 342.979\\
    Column density & \multirow{2}{3em}{1.20$\times$10$^{19}$} & \multirow{2}{3em}{2.35$\times$10$^{17}$} & \multirow{2}{3em}{1.20$\times$10$^{16}$}\\
    (cm$^{-2}$) & & & \\
    FWHM/2 (km s$^{-1}$) & 100 & 55.95 & 55.95\\
    $\tau$ & 8.30 & 5.94 & 3.44\\\hline
    \end{tabular}
    \tablefoot{Excitation temperatures were measured as the peak brightness temperature of the spectral feature. The values used for both isotopologues of SiO, such as temperature and partition function, are the same, with only column density varying between the calculations.}
    \label{table: tau}
\end{table}
The determined values of $\tau$ indicate optically thick emission. However, in the wings of the lines, which probe the outflowing gas shown in Fig. \ref{fig: moment maps}, the optical depth is considerably lower compared to the values given in Table \ref{table: tau}. Therefore, the optically thin wings of the lines represent well the source structure.
\section{XSHOOTER}\label{section: xshooter}
Here, we describe optical observations of V1309 Sco taken in 2016 and 2022.
\subsection{Observations}\label{section: xshooter obs}
\subsubsection{2016} \label{section: xshooter 2016 obs}
V1309 Sco was observed in 2016 using XSHOOTER \citep{vernet2011}, a mid-resolution spectrometer on UT 2 of the Very Large Telescope (VLT) which allows spectra to be taken simultaneously across three arms (UVB, VIS, NIR) with a total spectral coverage of 3000--24800 \AA. The spectra were acquired in stare mode. The slit width was 1\farcs3 for the UVB arm and 0\farcs9 for the VIS and NIR arms. The pixel binning for all arms was 1$\times$1. The data was observed in four separate runs. Run A was executed on 20$^{th}$ May 2016, runs B and C on 23$^{rd}$ June 2016, and run D on 6$^{th}$ July 2016. During the first run the telescope was incorrectly centered on a background star, and so run A was repeated on 6$^{th}$ July 2016.

In total, 4 exposures were acquired in the UVB and VIS arms and 20 exposures for NIR. The exposure times used were 2935, 2840, and 600 s for the UVB, VIS and NIR arms respectively.
\begin{table*}
\centering
\caption{Summary of XSHOOTER observations of V1309 Sco.}
    \begin{tabular}{c|cc|cc|cc}\hline
    & \multicolumn{2}{c|}{UVB} & \multicolumn{2}{c|}{VIS} & \multicolumn{2}{c}{NIR} \\
UT date & Exposure & Slit size & Exposure & Slit size & Exposure & Slit size \\
& time (s) & (\arcsec) & time (s) & (\arcsec) & time (s) & (\arcsec) \\\hline
20-05-2016 & 2935 & 1.3$\times$11 & 2840 & 0.9$\times$11 & 600$\times$5 & 0.9$\times$11 \\
23-06-2016 & 2935$\times$2 & 1.3$\times$11 & 2840$\times$2 & 0.9$\times$11 & 600$\times$10 & 0.9$\times$11 \\
06-07-2016 & 2935 & 1.3$\times$11 & 2840 & 0.9$\times$11 & 600$\times$5 & 0.9$\times$11 \\
26-10-2022 & 915$\times$3 & 1.3$\times$11 & 825$\times$3 & 0.9$\times$11 & 196$\times$14 & 0.6$\times$11 \\\hline
    \end{tabular}
    \label{table: xshooter obs}
\end{table*}
The data was reduced using XSHOOTER pipeline version 3.6.1 \citep{xshooterpipeline} in the \texttt{Esoreflex} environment, and further processed using standard IRAF routines. The data was calibrated using several standard stars: Hip084982, HD190285, Hip076069, Hip067973, Hip094378, Hip093049, Hip11900, Hip08254, and Hip017734. Dereddening was performed assuming $E(B$--$V)$=0.8 mag \citep{kaminski2015}. Table \ref{table: xshooter obs} presents a summary of the XSHOOTER observations.
\subsubsection{2022}\label{section: xshooter 2022 obs}
XSHOOTER observed V1309 Sco again on 26$^{\textrm{th}}$ October 2022, using a similar spectral setup to that used in 2016. However, in 2022, a narrower slit of 0\farcs6 was used for the NIR spectral arm, as the NIR emission was expected to have increased from the previous observation epoch. The exposure times were reduced, and more exposures in total taken. The total exposure time in each arm in 2022 is therefore significantly less than in 2016. The spectra were reduced and combined using the same pipeline version and method described in Sect. \ref{section: xshooter 2016 obs}. The standard star used to calibrate the spectra was Hip110573.
\subsection{Line identification and analysis}\label{section: xshooter line identification}
The XSHOOTER spectrum from 2016 is dominated by atomic and molecular emission. Atomic lines were identified by comparison to spectra in other Galactic red novae \citep{kaminski2009,kaminski2015,tylenda2015v4332}. In order to confirm uncertain lines, the NIST database\footnote{https://physics.nist.gov/PhysRefData/ASD/lines-form.html} was used to look for lines of the same multiplet, as well as compare relative intensities. Identified atomic lines are listed in Table \ref{table: xshooter lines}.

Examining the presence of molecules in the spectra, several oxygen-bearing molecules are identified: AlO, CrO, ScO, TiO, and VO. Multiple electronic systems have been identified across the three spectral arms.

Flux and width measurements of the 2016 observations were done using \texttt{splot}, as part of the \texttt{onedspec} package in IRAF. Errors are estimated at 5$\sigma$, where 1$\sigma$ is equal to the noise rms, and multiplying by the square root of the width of the line in pixels. Sections \ref{section: xshooter UVB lines}--\ref{section: xshooter NIR lines} describe in detail the detected atomic and molecular emission from each XSHOOTER arm in the 2016 and 2022 spectra.
\subsubsection{UVB range}\label{section: xshooter UVB lines}
The signal-to-noise ratio (SNR) of the UVB spectral range is lower than for the VIS and NIR ranges. Despite this, we still identified 26 individual atomic emission lines of different species, including S, Fe, Ca, and Cr. Amongst the strongest lines identified were those of Ca I $\lambda$4226, Cr I $\lambda$4254, and Fe I $\lambda$4375. The H$\beta$ emission was weaker than expected relative to other lines, but since H$\beta$ is partially blended with the AlO B$^2\Sigma^+$--X$^2\Sigma^+$ (1,1) band, the flux measurement on H$\beta$ is not reliable. Other significant features are the [S II] and [Fe II] lines, which are valuable due to their sensitivity to electron density and temperature. The Mn I triplet at $\sim$4028 \AA\space and the Cr I doublet at $\sim$5295 \AA\space could not be deblended and so flux measurements in Table \ref{table: xshooter lines} are presented as integrated flux measurements across the blended features. These measurements are shown for completeness and were not used in any detailed analysis.

The UVB range features molecular emission from all observed molecules except ScO, although less dominant than in the VIS and NIR spectra. For AlO and CrO, we see the B--X electronic systems; for VO we see C--X; TiO is traced through the $\alpha$ band emission. All such features at $\lambda >$ 5000 \AA\space could be modelled via the ExoMol--ExoCross\footnote{https://github.com/ExoMol/ExoCross} tool \citep{yurchenko2018}, whereas very few molecular bands were reliably identified at shorter wavelengths (see Fig. \ref{fig: molec models page 1}). As seen in Table \ref{table: molecular bands}, very few molecular bands were identified in the UVB spectrum.

Although there is a weak continuum seen in the UVB range, it is not strong enough to derive the spectral type and so is not analysed.
\subsubsection{VIS range}\label{section: xshooter VIS lines}
The VIS range has a similar number of atomic lines identified to UVB with 25 lines, although 6 are uncertain. One such line, found at a peak wavelength of 8662.33 \AA, was originally identified as Ca II $\lambda$8662 but was removed due to the inconsistent peak velocity with nearby atomic lines and other identified Ca II lines. The reasons for ambiguity in identification were mostly due to the observed central wavelengths not reliably matching the velocity measured for the majority of the lines, as well as due to possible contamination by molecular emission. For example, the unidentified line at 8601 \AA\space is heavily contaminated by the broad VO B$^4\Pi$--X$^4\Sigma^-$ (0,1) band. 

We see multiple forbidden lines of [O I], [S II], [Ca II] and [Fe II] as well as lines of Na I, K I, and Rb I. The strongest lines within the VIS range is the K I $\lambda\lambda$7664, 7698 doublet. We also see the semi-forbidden [Ca I line at 6572 \AA\space and the Li I 6708 \AA\space line \citep[for a summary of lithium in red nova remnants, see][]{kaminski2023}. The forbidden lines are important tools for constraints on the properties of the CSE and so are analysed in detail in Sect. \ref{section: diagnostics}.

The [S II] $\lambda$6730 line is contaminated by the TiO $\gamma$ (1,0) F$_1$--F$_1$ and TiO $\gamma$ (2,1) F$_3$--F$_3$ bands. As [S II] $\lambda$6730 is important for the constraints of the electron temperature and density, an accurate flux measurement for this line is important. We used the Pgopher\footnote{https://pgopher.chm.bris.ac.uk/} tool \citep{western2017pgopher} to model the TiO emission across a short wavelength range spanning the [S II] $\lambda$6730 line as well as the TiO $\gamma$ (0,0) bands in order to remove the molecular contamination. TiO was modelled to best fit the TiO $\gamma$ (0,0) F$_2$--F$_2$ band, as it had sufficient SNR and was devoid of saturation effects, unlike the TiO $\gamma$ (0,0) F$_3$--F$_3$ band. The best-fitting temperature was 240 K, with a Gaussian smoothing kernel of 3.5\,\AA. Pgopher was used rather than ExoMol--ExoCross for convenience, but is less accurate as it is parameterised by a single (rotational) temperature. The resulting flux measurements of the recovered [S II] $\lambda$6730 line is shown in Table \ref{table: xshooter lines}.

The molecular emission is far richer in the VIS range than the UVB. We see emission from all molecules listed in Table \ref{table: molecular bands}, with multiple bands detected for several molecules. The VIS spectrum is the only band where we observe ScO (A--X band between 6000--6100 \AA). At $\lambda <$ 7300 \AA\space the molecular spectrum is dominated by TiO and CrO whereas at longer wavelengths, VO is dominant. The key detections in the VIS spectrum are the various bands of the CrO B--X system, which has only been observed in one other astrophysical source (V4332 Sgr, \citeauthor{tylenda2015v4332} \citeyear{tylenda2015v4332}). V1309 Sco and V4332 Sgr remain the only sources in which we detect CrO emission. Additionally, as in \citet{kaminski2015}, we detect emission from the CrO A$^{\prime}$--X band observed between 8350 and 9500 \AA\space(Table \ref{table: molecular bands}).

Unlike in early spectroscopic observations of the remnant \citep{kaminski2015}, there is no visible continuum in the VIS range. 

\subsubsection{NIR range}\label{section: xshooter NIR lines}
As in 2012 \citep{kaminski2015}, there are very few atomic features seen in the NIR range. Ten atomic emission lines are seen, made up of [S II] and 6 unidentified lines. Three lines between 10290 and 10310 \AA\space cannot be identified, despite being relatively strong. They do not appear to be sky lines due to the broad widths comparable to the widths of the detected lines.

The molecular emission in the NIR range is just as rich as that seen for the VIS range. The NIR spectrum is dominated by the AlO A--X electronic bands, whilst CrO A--X, VO A--X, and ro-vibrational emission of AlO (X--X) are also seen. The strongest observed molecular feature across all spectral arms in absolute flux units, observed between 12100 and 12800 \AA, is the blended feature of the AlO  $\varv$=9--0 ro-vibrational emission, AlO A$^2\Pi_{1/2}$--X$^2\Sigma^+$ (3,0), AlO A$^2\Pi_{3/2}$--X$^2\Sigma^+$ (5,1), and CrO A$^5\Sigma^+$--X$^5\Pi$ (0,0) bands. The ro-vibrational H$_2$ 1--0 S(1) and 1--0 Q(1) lines are detected near 2.1 and 2.4 $\mu$m, respectively.

\subsection{2022 spectrum}
The 2022 XSHOOTER spectrum shows several changes in the spectroscopic appearance of the V1309 Sco remnant. The 2019 and 2022 spectra are compared in Figs. \ref{fig: xshooter epoch comparison 1} and \ref{fig: xshooter epoch comparison 2}. A significant difference is the absence of many atomic lines, including several forbidden iron and sulphur lines. Many of the atomic lines have also decreased significantly in intensity. Some of the more notable absences are the Rb I $\lambda$7947 line, the [S II] multiplet expected between 10280--10400 \AA, and the [Fe II] $\lambda\lambda$5159,5262 doublet. Additionally, the V I line at 3980.5 \AA\space was observed in the UVB spectrum that had not been previously seen in 2016. The weakening of the optical spectrum can be attributed to increased extinction due to growing dust column densities, indicating continuing dust formation since 2019. Therefore, constraining the physical properties of the optical emitting region of the remnant is becoming increasingly difficult for V1309 Sco, and is not attempted using the 2022 observations.

\subsection{Atomic line diagnostics}\label{section: diagnostics}
V1309 Sco shows no photospheric signatures from the merger remnant due to heavy circumstellar extinction caused by dust. The only visible source of emission we can study is the surrounding CSE, the origins of which are uncertain. Could the CSE be made up of material ejected during the binary interaction phase preceding the merger, or of material carried out by winds or outflows associated with the coalescence? By using diagnostic line ratios, we aim to examine the physical structure of the remnant. 

In order to examine whether LTE conditions apply to the atomic emission region, we compared selected line ratios of Fe I to simulated intensity ratios from the NIST service. The NIST-simulated ratios are calculated in LTE conditions, using the Boltzmann distribution and the Saha equation.

\begin{figure}
    \centering
    \includegraphics[trim=70 40 70 40, width=\columnwidth]{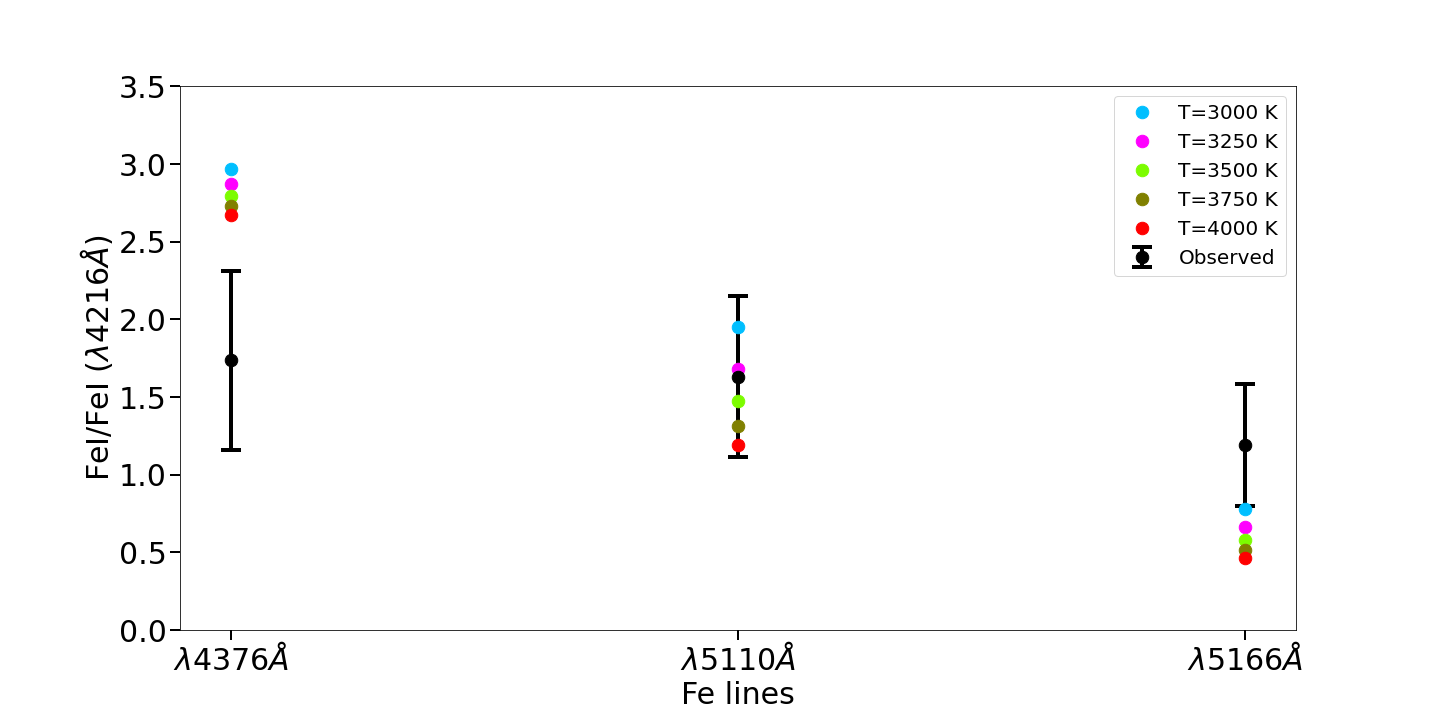}
    \caption{ NIST-simulated intensities of the Fe I lines at 4375, 5110 and 5166 \AA\space relative to the Fe I $\lambda$4216. The simulations were calculated at temperatures between  3000--4000 K, at intervals of 250 K. The $\chi^2$ test showed that the best fitting temperature was 3500 K (green points). The black points indicate the observed line ratios.}
    \label{fig: NIST}
\end{figure}
In Fig. \ref{fig: NIST}, we compare observed and simulated intensities of the detected Fe I lines relative to Fe I $\lambda$4216. The first set of simulations were extracted between 1000--10000 K at intervals of 1000 K. We estimated the best fitting electron temperature using a $\chi^2$ test on each model for the three line ratios, which gave a value of 3000 K. We then reduced the temperature range to 3000--4000 K with an interval of 250 K to refine the constraints. Figure \ref{fig: NIST} shows the results for the refined grid. The $\chi^2$ test yields a temperature of 3500 $\pm$ 250 K, with $\chi^2$=0.979. The poor consistency between the 3500 K model and the observed line ratios for the 4376 \AA\space and 5166 \AA\space lines may indicate a non-LTE environment.

Using PyNeb v1.1.16 \citep{luridiana2013, luridiana2015}, we attempted to get a more robust constraint on the electron temperature and density. The PyNeb tool makes no underlying assumptions about LTE, allowing the physical conditions to be constrained in non-LTE conditions. We constructed diagnostic maps similar to Figs. 2 and 3 in \citet{keenan1996}. For our diagnostics, we adopted the [S II] ratios from \citet{keenan1996} and the [Fe II] ratios from \citet{bautista2015}. For [S II], we used $\frac{4068}{6716+6731}$ vs $\frac{6716}{6731}$,  and replaced the 4068 \AA\space line with the other detected [S II] lines (except the [S II] $\lambda\lambda$6716,6731 doublet) to produce 5 different grids. For [Fe II], we plotted the $\frac{5159}{7155}$, $\frac{5262}{7155}$, $\frac{5159}{8617}$ and $\frac{5262}{8617}$ ratios against the $\frac{5159}{5262}$ and $\frac{7155}{8262}$ ratios. All line diagnostic grids are presented in logarithmic scale.

\begin{figure}
    \centering
    \includegraphics[width=0.5\textwidth, trim={2.4cm 3.5cm 4cm 5cm}, clip]{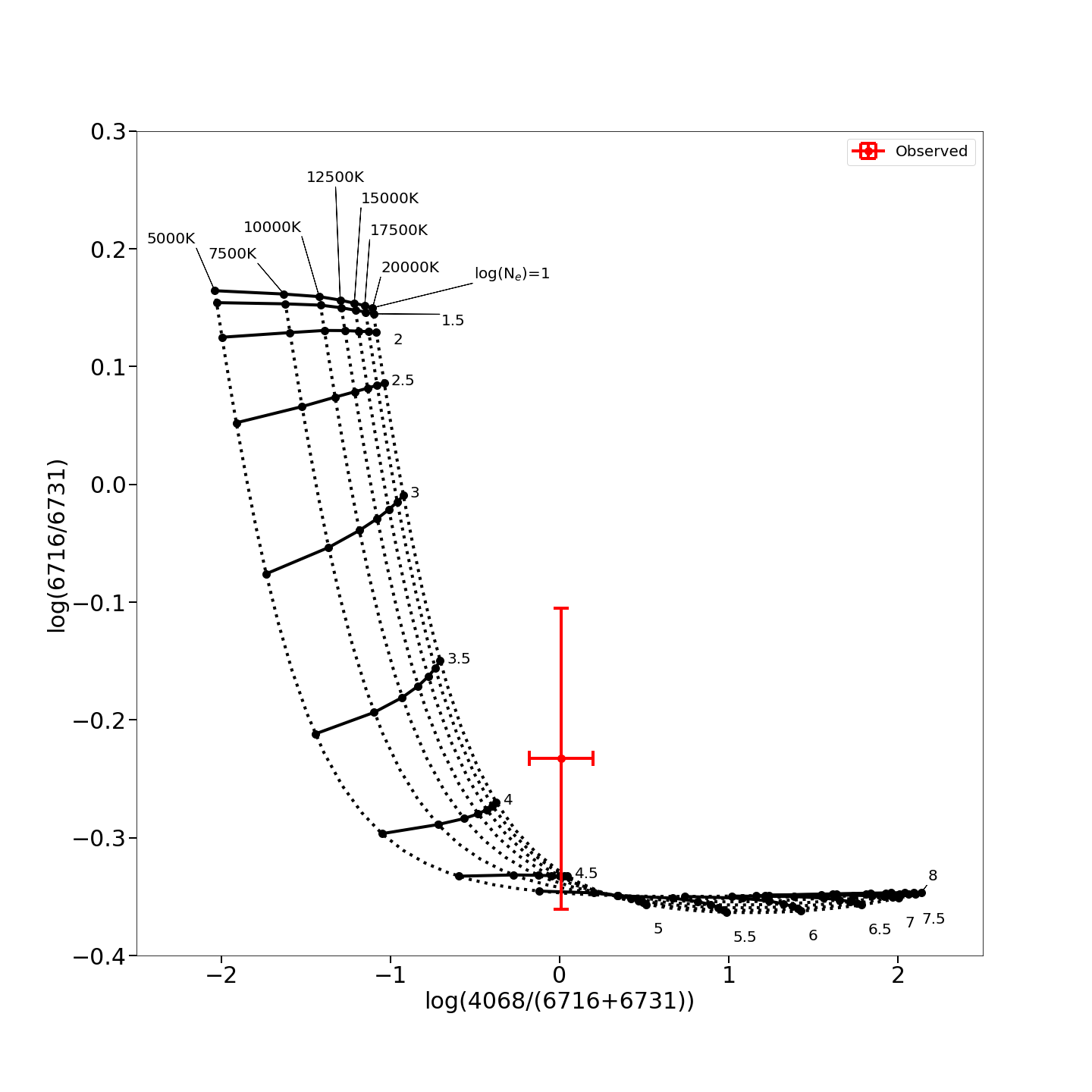}
    \caption{Diagnostic grid of [S II] $\frac{4068}{6716+6731}$ vs $\frac{6716}{6731}$. Solid lines represent lines of constant temperature and dotted lines represent lines of constant density. Larger black dots show the intersecting points between solid and dotted lines, and the red point and errorbars indicate the location of the observed quantity for both line ratios.}
    \label{fig: S2 4068 diagnostic}
\end{figure}

Figure \ref{fig: S2 4068 diagnostic} shows the [S II] $\frac{4068}{6716+6731}$ vs $\frac{6716}{6731}$ diagram. The location of the observed ratios on the grid suggests a lower limit on the electron density of $\log$($N_e$) $\leq$ 3.5. No constraint can be derived for the electron temperature $T_e$. We find that the grids involving the NIR [S II] lines are degenerate and therefore do not contribute to the constraints. Thus, little can be revealed about V1309 Sco by the [S II] diagnostics alone.

\begin{figure}
    \begin{center}
    \includegraphics[width=0.5\textwidth, trim={1.6cm 0 4cm 2cm}, clip]{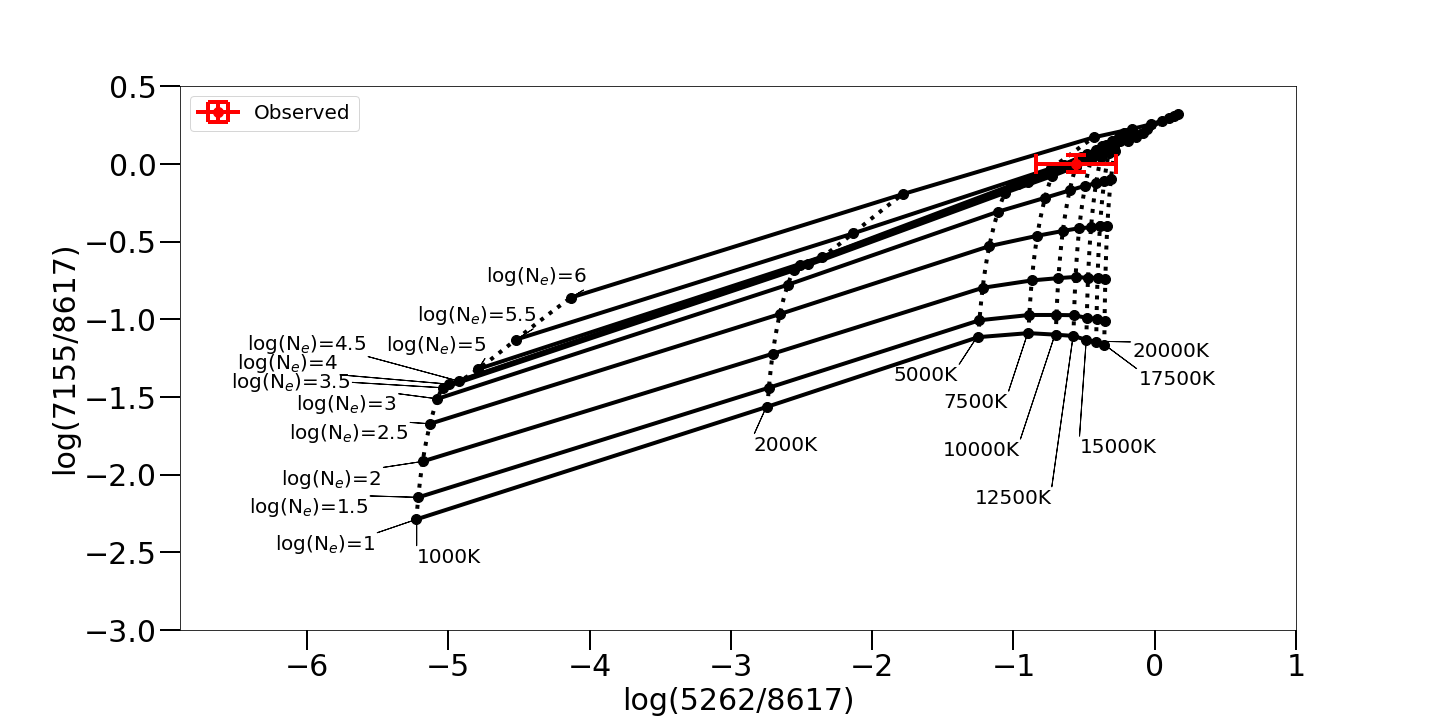}
    \includegraphics[width=0.5\textwidth, trim={1.6cm 0 4cm 2cm}, clip]{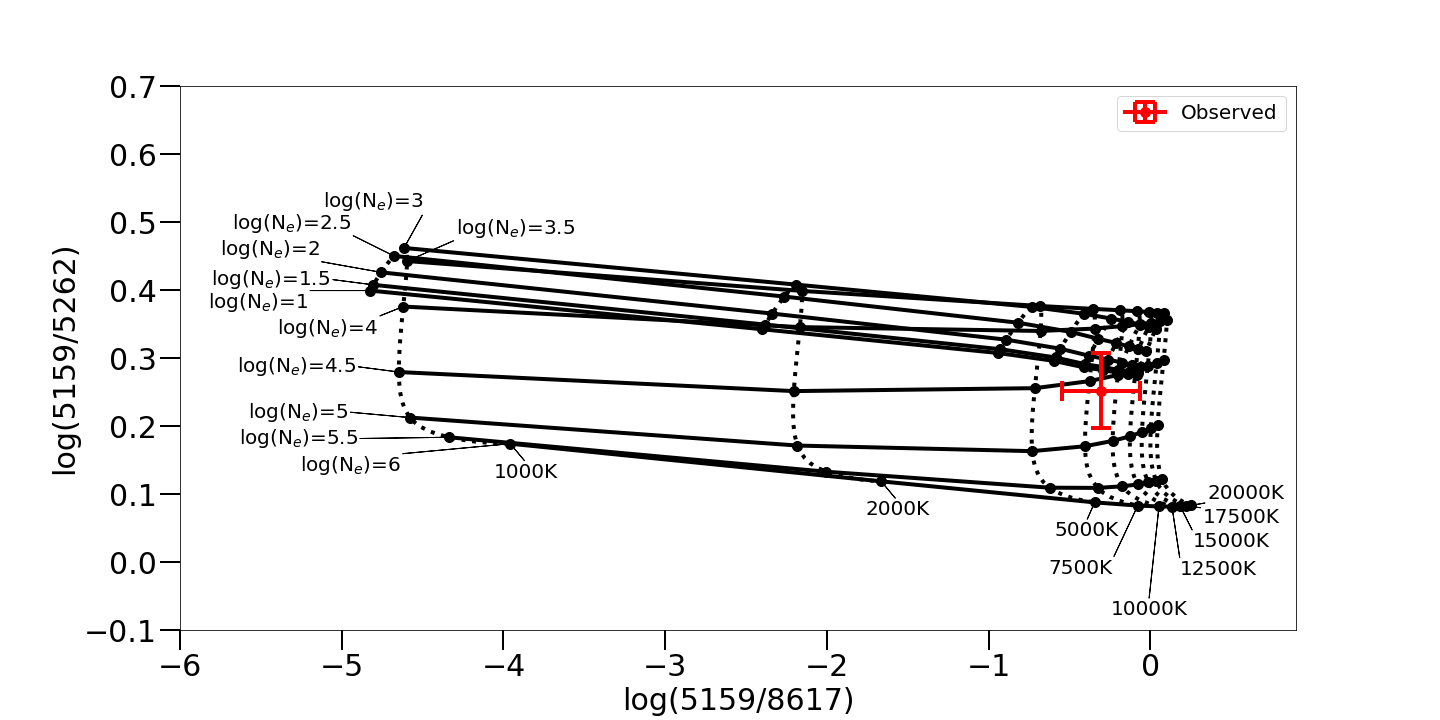}
    \includegraphics[width=0.5\textwidth, trim={2.4cm 0 4cm 2cm}, clip]{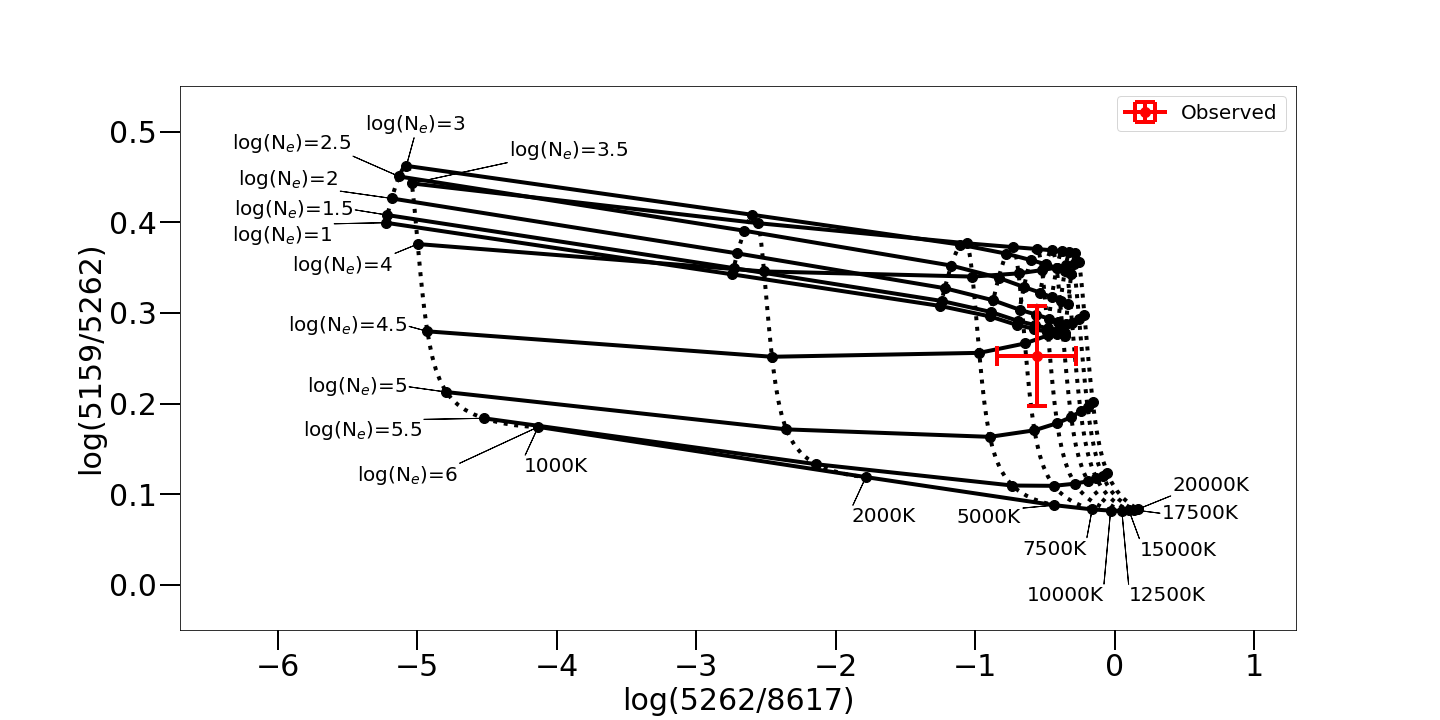}
    \caption{Same as Fig. \ref{fig: S2 4068 diagnostic}, but for [Fe II] $\frac{5262}{8617}$ vs $\frac{7155}{8617}$ (top panel), [Fe II] $\frac{5159}{8617}$ vs $\frac{5159}{5262}$ (middle panel) and $\frac{5262}{8617}$ vs $\frac{5159}{5262}$ (bottom panel).}
    \label{fig: Fe II diagnostics}
    \end{center}
\end{figure}

The top panel of Fig. \ref{fig: Fe II diagnostics} shows the diagnostic grid of [Fe II] $\frac{5262}{8617}$ vs $\frac{7155}{8617}$. The position of the observed ratios and associated errors are indicative of an $N_e$ range of 3 $\leq \log (N_e) \leq$ 5.5, as well as a lower $T_e$ limit of T$_e\space\geq$ 5000 K. The $N_e$ constraint inferred from Fig. \ref{fig: Fe II diagnostics} is consistent with the lower limit inferred by Fig. \ref{fig: S2 4068 diagnostic}. However, the constraints are loose as the parameter space is densely filled, and so constraints with lower uncertainty cannot be derived from [Fe II] $\frac{5262}{8617}$ vs $\frac{7155}{8617}$.

When we compare these constraints with the diagnostic grid of both [Fe II] $\frac{5159}{8617}$ vs $\frac{5159}{5262}$ (Fig. \ref{fig: Fe II diagnostics}, middle panel) and [Fe II] $\frac{5262}{8617}$ vs $\frac{5159}{5262}$ (Fig. \ref{fig: Fe II diagnostics}, bottom panel), the constraints can be tightened for both $N_e$ and $T_e$. The position of the observed line ratios and associated errors infer 4 $\leq \log (N_e) \leq$ 5 and 5000 $\leq$ $T_e\space\leq$ 15000 K, consistent with the observed quantities and diagnostic grids of [S II] $\frac{4068}{6716+6731}$ vs $\frac{6716}{6731}$ and [Fe II] $\frac{5262}{8617}$ vs $\frac{7155}{8617}$. However, as the 1$\sigma$ errorbars extend out between certain grid lines, we provide a more conservative constraint for $N_e$ to include the possibility that the errors are underestimated and due to the fact that the resolution of the diagnostic grids are quite low. The conservative constraint therefore is extended to 3 $\leq$ log($N_e$) $\leq$ 5.

As described in Sect. \ref{section: xshooter obs}, the data was corrected for interstellar reddening assuming $E(B$--$V)$=0.8 mag \citep{tylenda2011, kaminski2015}. In order to test this assumption, we used the python-implemented tool \texttt{extinction}\footnote{https://github.com/kbarbary/extinction}. We removed the previously applied extinction from the data and applied new corrections with different values of $E(B$--$V)$ to see if the constraints are affected. We re-reddened the data with $E(B$--$V)$=0.6 and 1.0 mag and plotted these values on the diagnostic grids. There was no significant effect of different $E(B$--$V)$ values on the diagnostics we derive.

To constrain the properties of shocks present in the CSE, we used the MAPPINGS V code \citep{mappings5}. We compared the measured integrated fluxes of atomic emission lines available in the MAPPINGS V model database to the models themselves. The models were primarily dependent on the gas density and shock velocity. Lines modelled in MAPPINGS include all the forbidden [S II] and [Fe II] lines, as well as [O I] and Fe I. However, the errors on the line fluxes proved too large to constrain either the gas density or shock velocity. Another issue was that for the densities we explored, which were dictated by the constraints on the gas density established from the PyNeb forbidden line diagnostics (Figs. \ref{fig: S2 4068 diagnostic}--\ref{fig: Fe II diagnostics}), the shock models were difficult to resolve at different densities for many of the modelled emission lines.

\subsection{Molecular modelling}\label{section: molecular modelling}
Using the ExoMol-ExoCross modelling described in Appendix \ref{appendix: xshooter molecules}, we attempted to constrain the rotational (T$_{rot}$) and vibrational temperatures (T$_{vib}$) of all detected molecules in our XSHOOTER spectra. The results are T$_{rot}$=200 and 300 K and T$_{vib}$=1700 and 2400 K for AlO and TiO, respectively. The fitting routine could not converge for CrO, ScO, and VO, and so we instead take the average ro-vibrational temperatures fitted for AlO and TiO, T$_{rot}$=200 K and T$_{vib}$=2100 K, and apply these to the other three detected molecules to obtain their simulated spectra. The overall model, as well as individual contributions from each molecule, is presented in figures of Appendix \ref{appendix: molecular models}.


\section{Discussion}\label{section: discussion}
 \subsection{CSE structure}\label{discussion: kinetic structure}
Very little theoretical work has been done on the post-merger evolution of red novae, with studies concentrating primarily on the merging process \citep[e.g.,][]{zhu2013,pejcha2014,pejcha2017,nandez2014, metzger2017, Iaconi2017, Iaconi2020, soker2023}. Therefore, any connections between theoretical and observational studies of red novae cannot be fully verified. Nevertheless, the angular momentum of the binary is expected to impose some form of bipolar structure in the remnant, but whether it could be observed at any stage and how much mass it carries are open issues.
 
Our ALMA maps only marginally resolve the cool component of the V1309 Sco remnant. The spatial structure is not easy to recognize at the achieved resolution, but a slight elongation can be seen in mapped molecular lines. Additionally, moment-1 maps show a velocity gradient along the same axis. These features are consistent with a bipolar configuration. A large part of the molecular gas must also be located near the systemic (average) velocity and near the central source to produce the centrally-peaked moment-0 maps shown with contours in Fig. \ref{fig: moment maps}. Such observational characteristics are consistent with what ALMA observed for V4332 Sgr, the older and larger clone of V1309 Sco (K18). Using ALMA data of V4332 Sgr, which better resolved the source, \textbf{K18} constructed a full 3-dimensional (plus kinematical) model of the remnant, which prominently shows a pair of bipolar lobes formed by opposite wide-angle outflows. We postulate that the remnant of V1309 Sco displays very similar architecture.



\begin{figure}
    \centering
    \includegraphics[width=0.5\textwidth, trim={1cm 0cm 2cm 3cm }, clip]{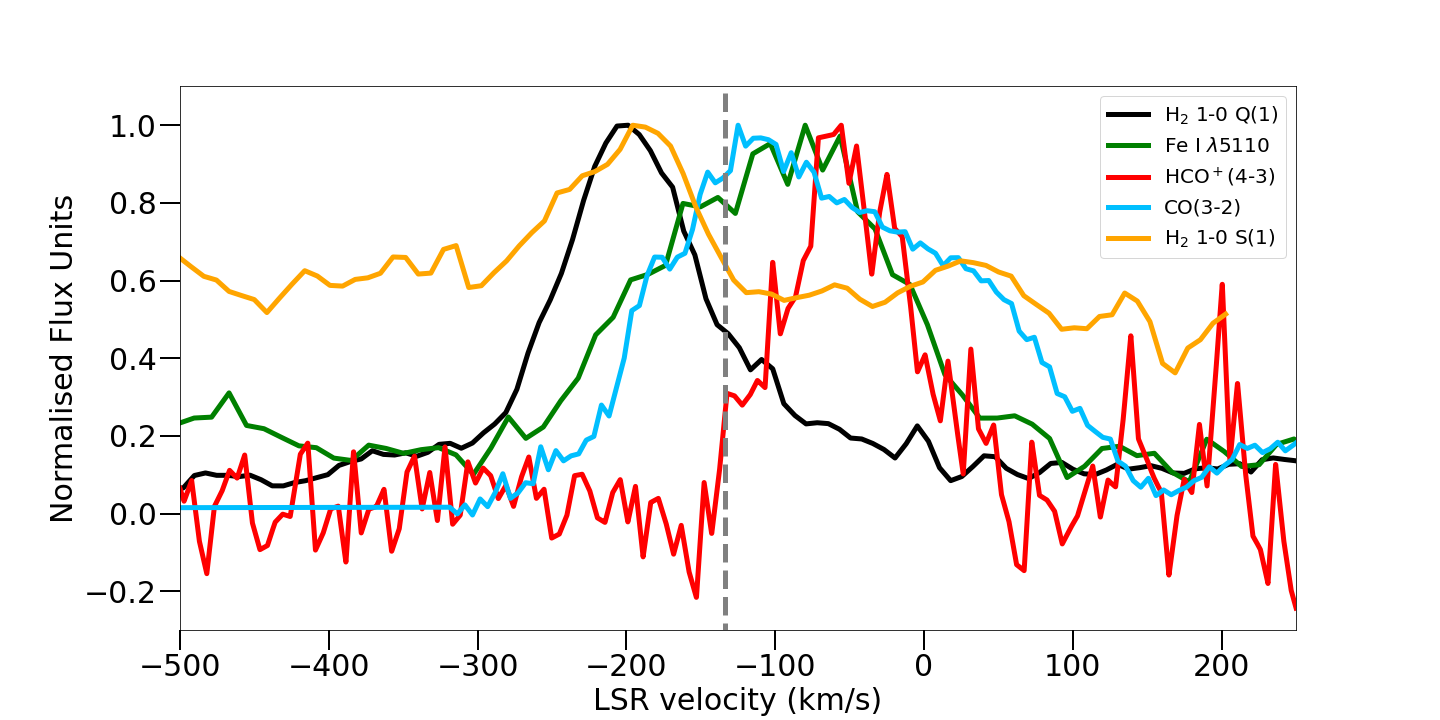}
    \includegraphics[width=0.5\textwidth, trim={1cm 0cm 2cm 2cm }, clip]{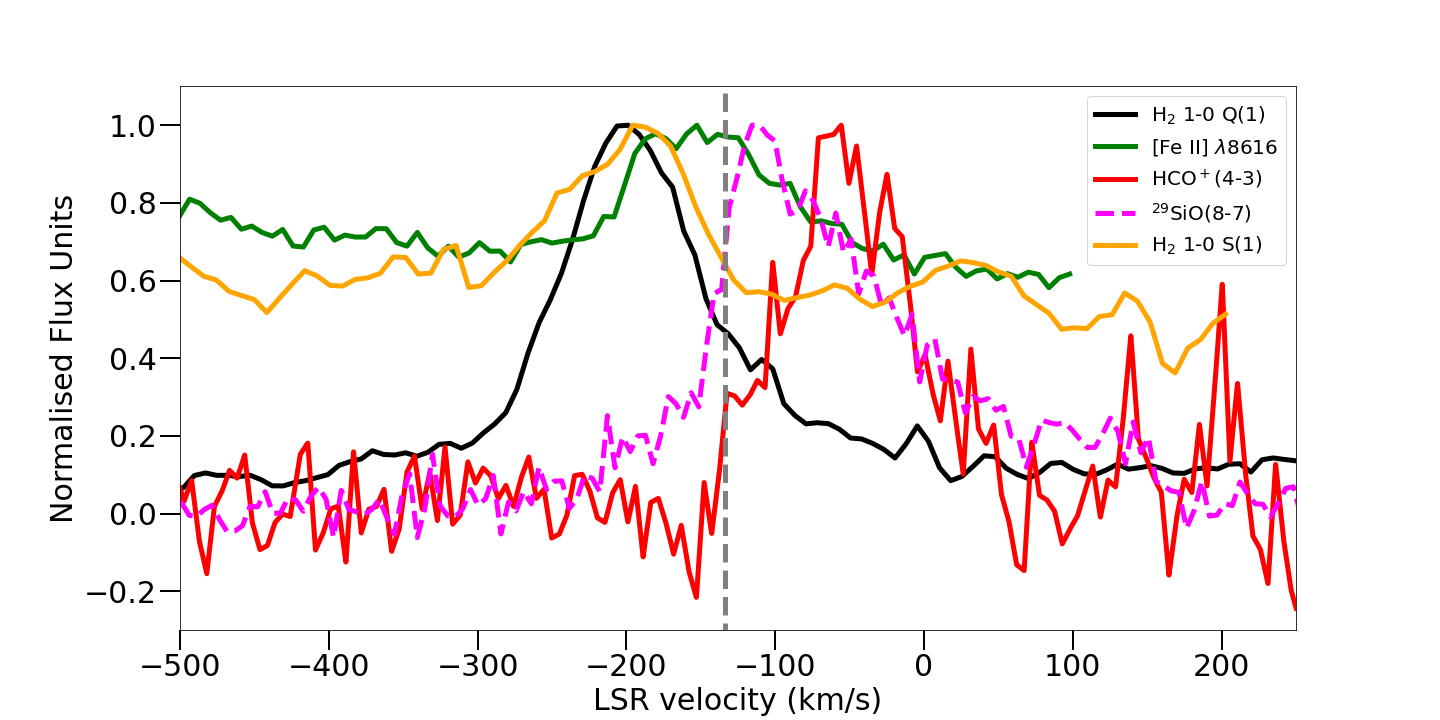}
    \caption{Velocity profiles of HCO$^+$ (4--3), H$_2$ 1--0 S(1) and H$_2$ 1--0 Q(1), normalised or otherwise scaled and shifted (both panels). The top panel also shows the CO (3--2) and Fe I $\lambda$5110 profiles, whilst the bottom panel shows [Fe II] $\lambda$8616 and $^{29}$SiO (8--7). The grey dashed line represents the average peak velocity of the atomic emission detected in XSHOOTER at --134 km s$^{-1}$.}
    \label{fig: HCO+ H2 comparison}
\end{figure}

This bipolar structure can be considered to be reflected in our radiative transfer modeling of the sub-mm lines, in which the emission features are represented by two kinematical components, NC and BC. Since these components are not readily separated in the spectra and overlap for all observed species, deciding on modelling two components is merely a simplification which was thought to produce more robust constraints on the gas properties than any single-component model. Fitting more than two components would require even more parameters, and is less informative. However, the observations may suggest that more than two kinematic components form the observed spectral profiles. Indeed, the line profiles of CO and $^{29}$SiO representing the entire source show an excess within the red wings of the lines that is not replicated in our CASSIS model (Fig. \ref{fig: LTE model}). This excess is also clearly seen in Fig. \ref{fig: HCO+ H2 comparison}. There is some asymmetry in the gas distribution, which we are not able to resolve with the current sub-mm data.


In our simplifying model, the two postulated lobes are represented by the narrow and broad Gaussian components, whose properties for both ALMA epochs are displayed in Table \ref{table: chi^2}. The automatic fit is not consistent in assigning the positions and widths of the components when the two epochs are compared. In the earlier epoch it is the blue component which is wider, whereas in the 2019 fit the blue component is narrower. It is unlikely to be a real change in the physical parameters of the flow and rather reflects the large uncertainties in modelling the multiple overlapping components. As discussed in Sec. \ref{section: ALMA tau}, the moderate to high optical depths of the mapped emission may be another source of uncertainty. A FWHM of 150\,km s$^{-1}$, representing an average of all simulated components, is probably the best educated guess on the actual (projected) velocity dispersion in both lobes.

The bulk of material in the CSE is likely represented by the molecular emission observed with ALMA. This material has cooled to temperatures of 35--113 K, and recombined to form molecules. Slightly warmer molecular gas at a temperature of 200 K is traced by the optical spectra presented in Appendix \ref{appendix: ALMA lines}. Here, we assume that the rotational temperature is close to the excitation temperature, whilst the vibrational temperature of 2100 K represents the colour temperature of the radiation field responsible for the fluorescent emission in the electronic molecular bands. Atomic gas traced by the same optical spectra exhibits much higher temperatures of 5--15 kK, revealing hot gas with partial ionisation.

In Fig. \ref{fig: HCO+ H2 comparison}, we compare the velocity profiles of Fe I and [Fe II], tracing the hot gas, with those of the sub-mm lines of CO and $^{29}$SiO which represents the coolest gas observed. The profiles overlap closely, although the sub-mm lines show stronger red wings relative to the optical lines at positive velocities. It appears that the kinematics of the hot and cold gas regions are similar within the remnant, but this does not mean that they are chemically mixed. In classical cool circumstellar envelopes (such as those surrounding AGB stars), warmer gas is usually found closer to the central star which heats the gas via radiation or shocks \citep{bell1993,olguin2020,massalkhi2020}. It is therefore likely that the atomic gas is much more compact than the sub-mm molecular gas. 

Therefore attempts to resolve the remnant are likely to be more successful using sub-mm observations rather than optical/IR. Better angular resolution is also required to resolve the overall structure of the V1309 Sco remnant.

\subsection{Role of shocks}\label{discussion: outflows}
In analogy with other Galactic red nova remnants \citep{kaminski2009v838mon,kaminski2010v4332,tylenda2015v4332}, the remnant star of V1309 Sco is expected to be a cool (2000--3000 K) giant or supergiant. Section \ref{discussion: kinetic structure} describes the detection of atomic gas with temperatures exceeding 5000 K. The presence of hot gas in the vicinity of a cool star would therefore be significant. The gas may be material that was heated during the 2008 eruption, cooling on a timescale of decades and recombining into molecular gas. However, the detection of NIR H$_2$ and sub-mm HCO$^+$ emission suggests that the material may be heated by active shocks propagating through the CSE, subsequently changing its molecular composition.
\subsubsection{HCO$^+$}
In Sect. \ref{section: ALMA line ident}, we postulated the identification of the HCO$^+$ $J$=4--3 line in the 2019 spectrum, which must have been much weaker or absent in the 2016 spectrum (see Fig. \ref{fig: ALMA spectrum labelled}). Although the HCO$^+$ line is heavily blended with SO$_2$, the radiative transfer model of SO$_2$ is good enough to extract the intrinsic line profile of HCO$^+$. This profile is shown in Fig. \ref{fig: HCO+ H2 comparison}.

The recovered HCO$^+$ profile has a peak velocity of --53 km s$^{-1}$, and its profile partially overlaps in velocity with the profiles of other sub-mm and optical lines. The HCO$^+$ emission is the most redshifted emission detected across both the optical and sub-mm. The shift between HCO$^+$ and the average velocity of the atomic gas is 81 km s$^{-1}$.

HCO$^+$ is often considered a shock tracer, especially in cooler environments. Shock dissociation of stable molecules such as H$_2$O and CO, and subsequent recombination can form HCO$^+$ in circumstellar media \citep{pulliam2011, sanchez1997, sanchez2000, sanchez2015}. Alternatively, ultraviolet and X-ray fluxes can influence HCO$^+$ abundances through photo-dissociation \citep{kimura2012,cleeves2017}. Due to a lack of UV or X-ray sources in V1309 Sco, shocks are the most likely candidate for the formation of HCO$^+$.

\subsubsection{H$_2$}
Ro-vibrational 1--0 S(1) and Q(1) lines of H$_2$ are detected in our XSHOOTER NIR spectrum at 2.10 and 2.41 $\mu$m, respectively. Ro-vibrational H$_2$ lines have long been considered a shock tracer, and have often been used in conjunction with NIR [Fe II] to probe different regions of the same shock \citep{sternberg1989,davis2003,kokusho2020,mohan2023}. We detect no such NIR [Fe II] in our XSHOOTER spectrum. In Fig. \ref{fig: HCO+ H2 comparison}, we plot both detected H$_2$ line profiles compared to sub-mm and optical lines. The line profiles of each line are found at the same peak LSR velocity of --216 km s$^{-1}$, and partly overlap with the velocity profiles of CO, Fe I and [Fe II]. The blue wings of both H$_2$ lines do not overlap at all with other observed species. The shift between the average atomic gas velocity and the velocity of the H$_2$ lines is 82 km s$^{-1}$.

In order to derive shock parameters from H$_2$, we examine the H$_2$ shock model grid presented in \citet{kristensen2023}. The grid is calculated using the Paris-Dunham code \citep{godard2019}, covering six parameters; pre-shock density, shock velocity, transverse magnetic field strength, external UV radiation, H$_2$ cosmic-ray ionisation rate and fractional abundance of polycyclic aromatic hydrocarbons (PAHs). The authors identify the twenty-five dominant cooling lines, eight of which are covered in our XSHOOTER spectra. Of those eight, three lines lie within the telluric absorption bands (see Fig. \ref{fig: molec models page 2}) and so are unreliable in estimating upper limits of fluxes. Three more lines, 1--0 Q(3), 1--0 S(2), and 1--0 Q(5) are not detected. The 1--0 S(2) line is heavily obscured by AlO, and so we measure the flux of the AlO feature and apply the integrated flux as a conservative upper limit. The other two lines are not detected above rms noise levels. We therefore simulated the fluxes of a Gaussian with peak flux equal to the rms noise in the region where we expect the line to be detected, with the same FWHM as the average of the two detected lines. Using the two detected and three undetected lines, we calculate the observed line ratios of the 1--0 S(1) line and three upper limits against 1--0 Q(1), the stronger of the two detected lines. We then search through the model grid to find models which are in agreement with the observed ratios to 5$\sigma$ accuracy for the detected line ratio (where $\sigma$ is the line ratio error calculated using standard quadrature equations), and are below the ratio upper limits calculated for the simulated lines. 

\citet{nicholls2013} find no evidence of PAHs in their analysis of the mid-IR SED of V1309 Sco. Therefore, we only examine models with the lowest fractional abundance of PAHs modelled (=10$^{-8}$). However, we find it has no real effect on the statistics presented in Appendix \ref{appendix: H2 models}.
Using the conditions described above, we find that 139 out of 14364 models are in agreement. We present statistics of the consistent models in Appendix \ref{appendix: H2 models}.

We find some initial constraints on the shocks related to the H$_2$ emission (Fig. \ref{fig: H2 models}), although the external UV radiation and H$_2$ cosmic-ray ionisation rate cannot be constrained. The models indicate a pre-shock density lower limit of 10$^7$ cm$^{-3}$. This is a lower limit as larger densities are not covered by the models, but may still be the true pre-shock density. The scaled value of the magnetic field covers 0.1 to 1.0 (while the grid spans to values as high as 10). For the shock velocity, we see two separate subsets of models at lower and higher shock velocities (Fig. \ref{fig: H2 models}, third panel). Models with low (v$_s<$10 km s$^{-1}$) shock velocities are typically found at lower magnetic fields, whereas higher shock velocities (20--30 km s$^{-1}$) are found across a wide range of magnetic field strengths (Fig. \ref{fig: H2 high+low vel}). Finally, almost all models fulfilling our observational constraints exhibit J-type shocks.

\subsubsection{The properties of the shocks}
\begin{figure}
    \centering
    \includegraphics[width=0.5\textwidth]{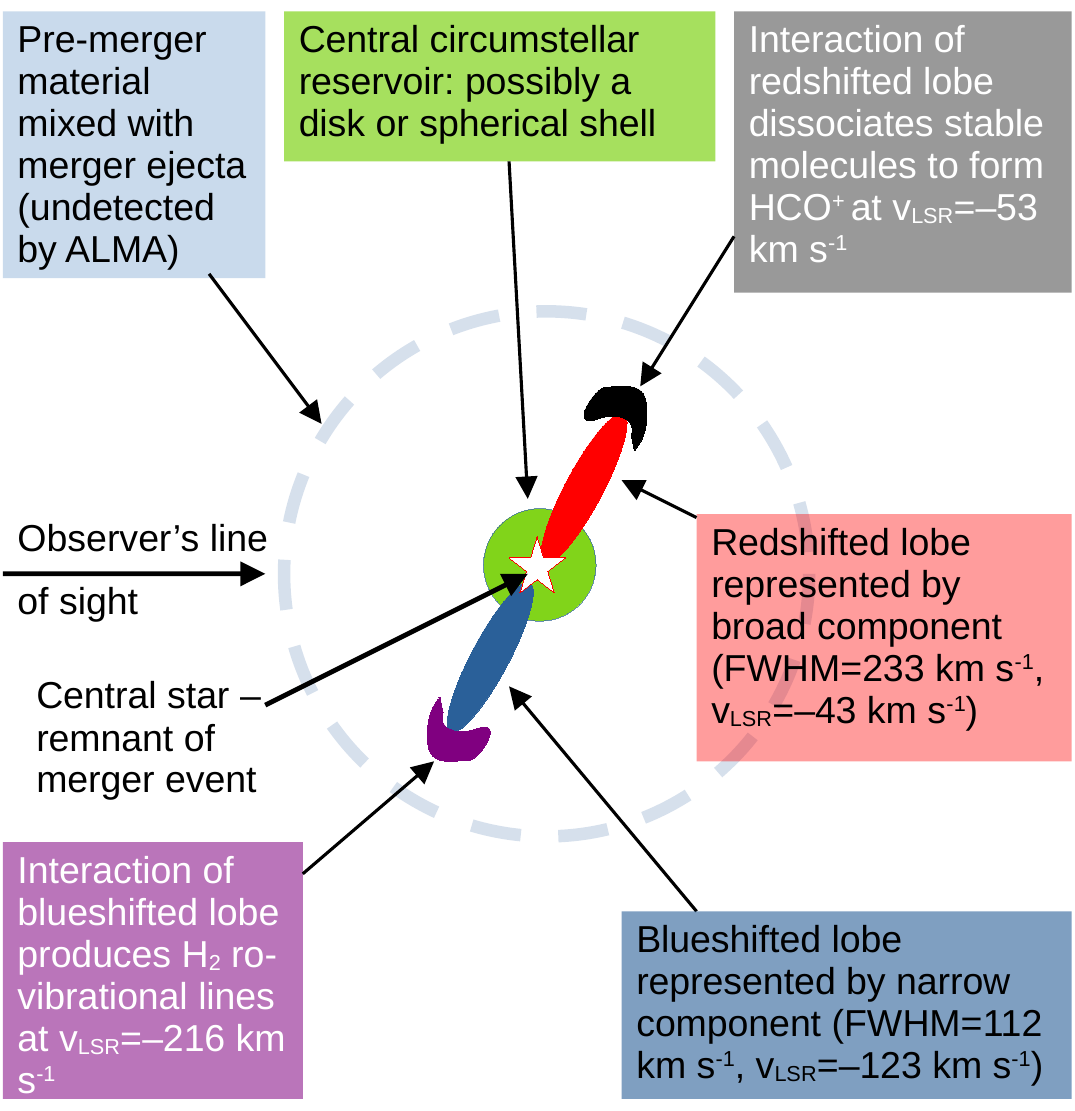}
    \caption{Schematic of the bipolar structure hypothesised in V1309 Sco. The velocities for the blueshifted and redshifted lobes are measured from the 2019 ALMA data.}
    \label{fig: schematic}
\end{figure}

We have proposed that both NIR H$_2$ and sub-mm HCO$^+$ emission probes shock-excited regions in V1309 Sco, but -- as shown in Fig. \ref{fig: HCO+ H2 comparison} -- these shock tracers have very different kinematics with H$_2$ observed at extreme blueshifted flow and HCO$^+$ in the extreme redshifted flow. 
Fig. \ref{fig: HCO+ H2 comparison} shows an almost equal magnitude in the velocity shift of the shock tracers ($\sim$80 km s$^{-1}$) relative to the average velocity of the atomic emission. This symmetric location can be explained if the emission arises in two opposite flows, as schematically illustrated in Fig. \ref{fig: schematic}. Since the fastest ejected gas is expected farthest from the central object, we place the shock regions at the apexes of the bipolar flows, which are seen in cool molecular gas observed with ALMA. However, it is unclear why the blueshifted lobe produces only ro-vibrational H$_2$ emission, whereas we only see redshifted HCO$^+$ emission. This suggests that the shocks propagating in opposite directions interact with the ambient circumstellar gas in very different ways. This could depend on either the properties of the ambient gas into which the shocks are propagating, or even the shocks themselves. It could also be that the NIR H$_2$ emission excited by shocks via the redshifted lobe is compact and could be obscured by circumstellar dust. However, this does not explain why we do not see blueshifted HCO$^+$ emission. The sub-mm emission is not affected by dust extinction.

The results presented in Appendix \ref{appendix: H2 models} show that the shocks that excite H$_2$ could be classified as J-type shocks. The results also suggest that the shock velocities lie in one of two regions, in the range of 2--10 km s$^{-1}$ or are faster than 15 km s$^{-1}$. High-velocity J-type shocks with shock velocities greater than 20 km s$^{-1}$ would be expected to dissociate H$_2$ \citep[][and references therein]{flower2003}, but this may not be the case for lower velocity shocks. If the shocks associated with HCO$^+$ do have higher shock velocities, the absence of ro-vibrational H$_2$ in the HCO$^+$ excitation region may be due to H$_2$ being efficiently dissociated. On the other hand, if the shocks propagating via the blueshifted lobe had lower shock velocities, it is possible that the shocks have sufficient energy to excite H$_2$, but not enough to dissociate stable molecules such as CO that would go on to form HCO$^+$. 

An alternative explanation would be that there is an increased column density of CO surrounding the redshifted lobe that could lead to increased HCO$^+$ formation. Indeed, as mentioned, The CO 3--2 line profile does show excess emission in the red wing.

\subsection{Effects of dust obscuration}\label{discussion: line profiles}
In the XSHOOTER spectra, we detect a plethora of neutral and singly-ionised emission, as well as forbidden line emission (see Sect. \ref{section: diagnostics}). It has been shown that the line profiles of atomic emission in core-collapse supernovae is affected by dust \citep[CCSNe][]{lucy1989,pozzo2004,bevan2017}, with a bluewards shift in the peak velocity and asymmetries between the red and blue wings. \citet{shore2018} applied a similar method to the ejecta of classical novae, which have simpler geometries and lower ejecta velocities and masses. Similar effects are seen, as well as that in cases where the inclination is low so that the line of sight barely passes through the ejected material, asymmetrries will still be seen. To examine the effect of dust on the neutral and ionised atomic emission, we looked at the profiles of neutral and singly ionised Fe and Ca. The profiles are shown in Fig. \ref{fig: Ca, Fe profiles}. The average profiles of [Ca II] and [Fe II] in 2016 show a slight asymmetry skewed towards shorter wavelengths, with the peak velocity also blueshifted. This would indicate that ionised gas is more obscured by dust and so is likely to be found at smaller radial distances from the central star, assuming the winds or outflows are biconical and sufficiently inclined to the observer that both cones are visible. Such a geometry is consistent with the ALMA velocity maps shown in Fig. \ref{fig: moment maps}, which in 2019 are similar to that seen for V4332 Sgr \citep{kaminski2018}. In \textbf{K18}, it is noted that the inclination angle may be even lower than that found for V4332 Sgr (13$\degree$). If this is the case, the assumed geometry in the simulations of \citeauthor{shore2018} can be applied to V1309 Sco.
\begin{figure}
    \centering
    \includegraphics[width=0.5\textwidth, trim={0 3.2cm 0 5.5cm}, clip]{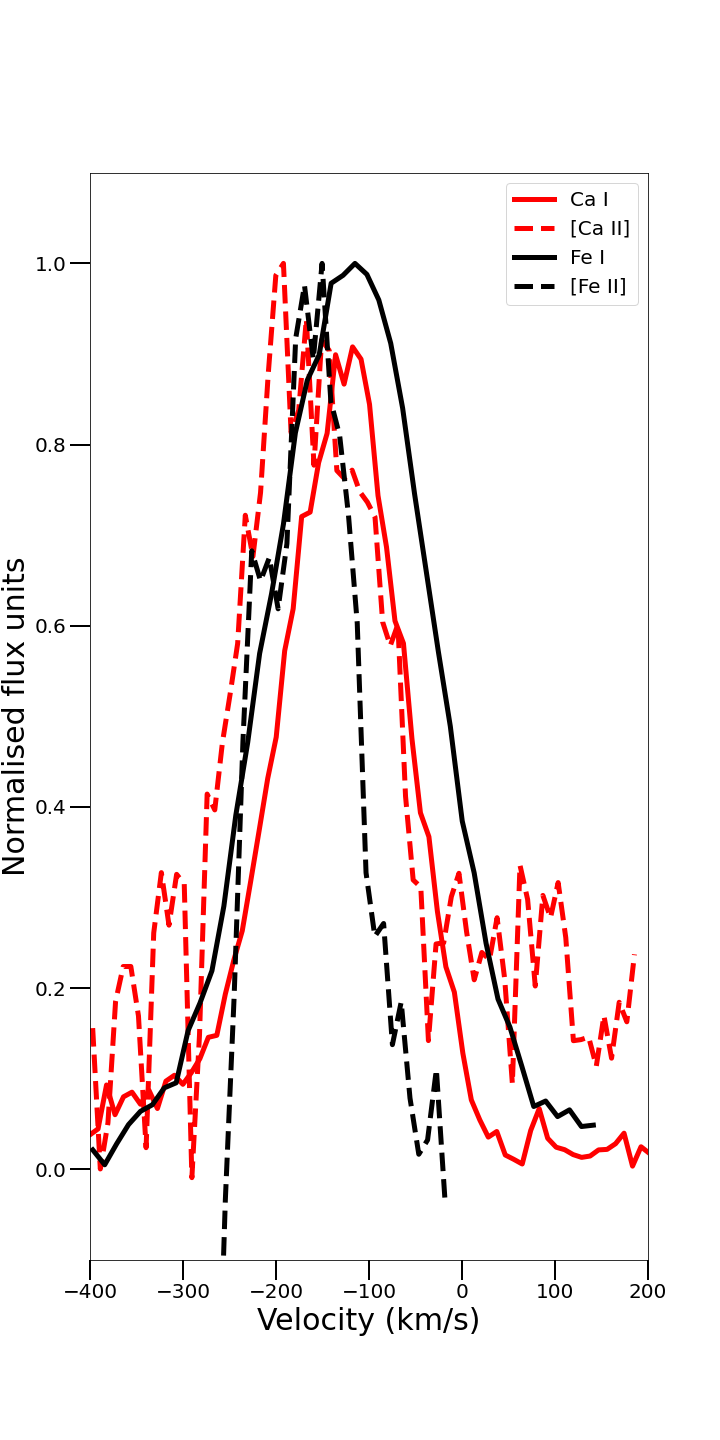}
    \caption{Averaged profiles of neutral and singly ionised Ca and Fe. Black and red lines show Fe and Ca respectively, whilst solid and dashed lines show neutral and singly ionised line profiles. The profiles of ionised species show asymmetry and a shift towards bluer velocities, possibly indicating greater obscuration by dust.}
    \label{fig: Ca, Fe profiles}
\end{figure}
\begin{figure}[h!]
    \centering
    \includegraphics[width=0.5\textwidth, trim={2cm 0 0 2cm}, clip]{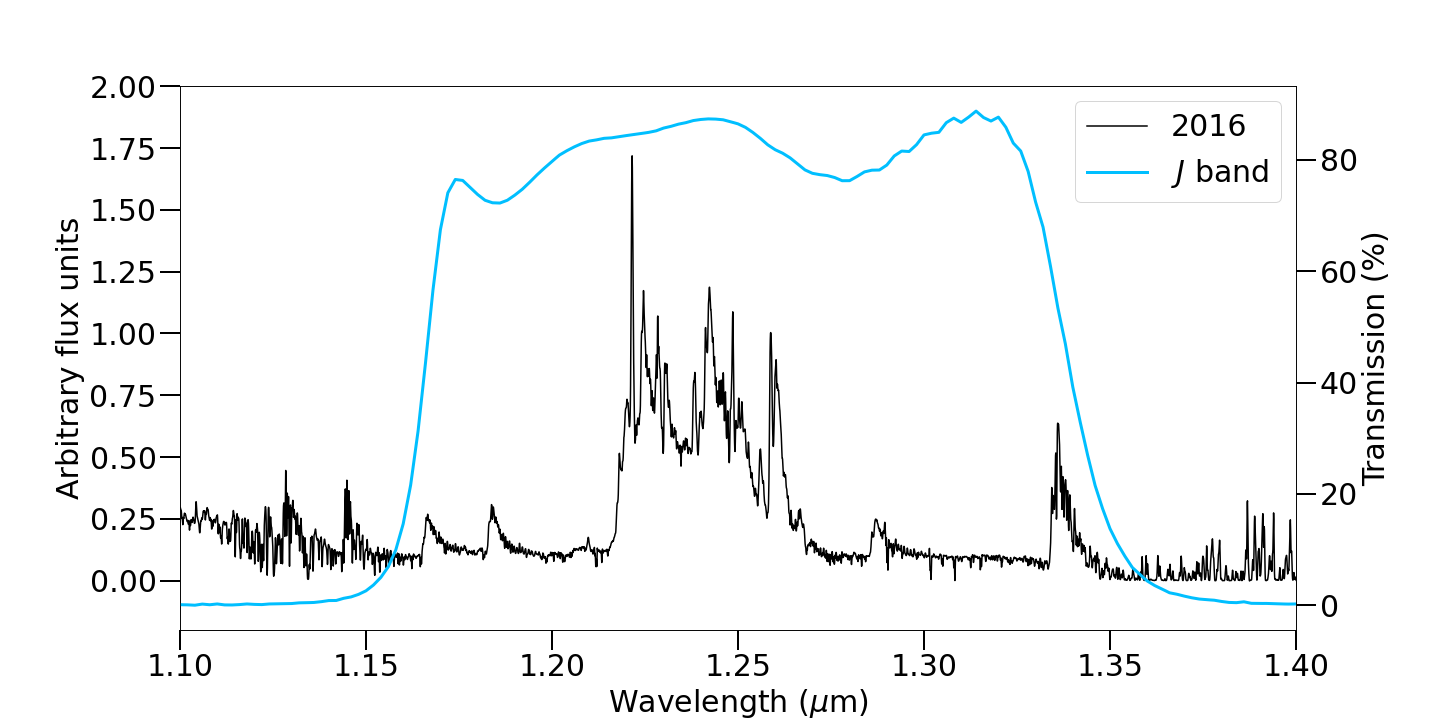}
    \includegraphics[width=0.5\textwidth, trim={2cm 0 0 2cm}, clip]{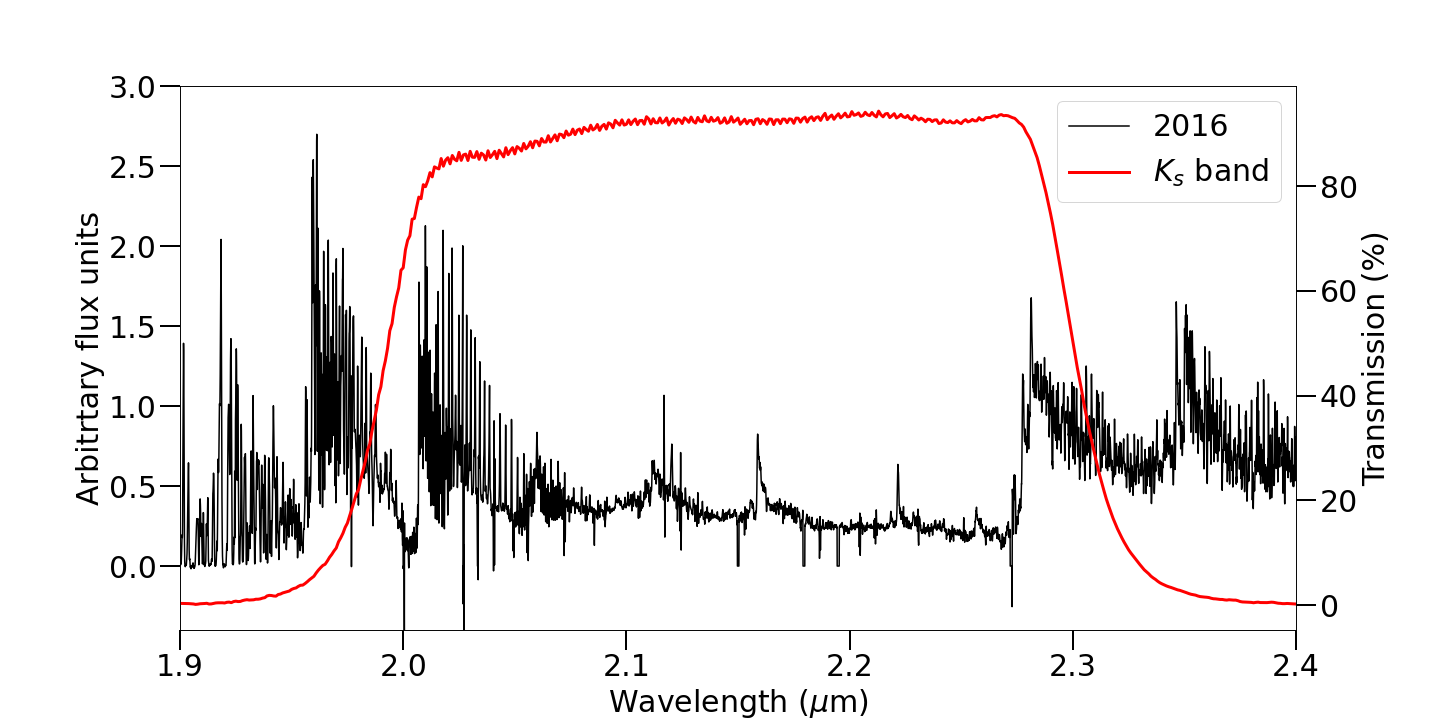}
    \caption{\textbf{Top}: The XSHOOTER spectra between 11300--14100\AA\space from 2016 (black) with the transmission curve of the $J$ band used in the VVV survey overplotted (blue). \textbf{Bottom:} Same as top, but between 19000--24100 \AA\space and with the $K_s$ band overplotted (red).}
    \label{fig: NIR bands}
\end{figure}
\subsection{V1309 Sco is not a blue straggler}\label{discussion: blue straggler}
The VISTA Variables in V\'ia Lactea \citep[VVV][]{minniti2010VVV} survey observed V1309 Sco across multiple epochs between 2010 and 2015. Analysis by \citet{ferreira2019} showed that the near-infrared $J-K_s$ colour decreased from 1.40 to 0.42 mag between 2010 and 2015. In addition, the $I-K_s$ colour \citep[where $I$ is the OGLE $I$-band, see][]{tylenda2011} changed from 3.54 mag in 2010 to 2.75 mag in 2015. \citeauthor{ferreira2019} use the apparent shift towards bluer colours between 2010 and 2015, as well as the asymptotic decline in the $K_s$ band, to conclude that V1309 Sco is a blue straggler \citep{sandage1953}. The \textit{J} band covers a wavelength range of 1--1.4 $\mu$m, and the $K_s$ band covers 1.8--2.6 $\mu$m. Our 2016 XSHOOTER spectra presented in Fig. \ref{fig: NIR bands} in those spectral ranges show that the $J$ filter is dominated by a prominent molecular band featuring emission from AlO and CrO. \citet{kaminski2015} note that in 2009, V1309 Sco spectra showed molecular absorption bands which diminished towards the end of the year. By 2012, much of the molecular absorption had transitioned to pure emission in their spectra. It seems as though the 2010 VVV observations were taken during this transition period, with the 2015 observations taken after the molecular emission had been established. As the $J$ band is dominated by molecular emission rather than continuum, the apparent increase in blue luminosity in the near-infrared is attributed to the emergence of molecular emission rather than the evolution of V1309 Sco to a blue straggler.
Regarding the $I$--$K_s$ colour evolution, the $K_s$ band is less dominated by molecular emission, although some AlO A--X emission is seen at the edge of the spectral coverage. As seen in Fig. \ref{fig: NIR bands}, no continuum is detected within the $K_s$ band. Our XSHOOTER observations are much more sensitive and yet no continuum is detected that can be attributed to V1309 Sco. V1309 Sco is located in a crowded field, and so it is possible for mistakenly associating continuum emission with V1309 Sco. \citet{kaminski2015} note that the continuum source disappears between 2009 and 2012, meaning that the evolution of the $I$--$K_s$ colour can be attributed to a disappearance of continuum entirely. We therefore believe that V1309 Sco is not evolving towards a blue straggler.

\section{Summary}\label{section: conclusions}
Using optical spectroscopy from 2016 and sub-mm interferometry from 2019, we examine the circumstellar environment of the stellar merger remnant V1309 Sco in order to understand the kinematical structure of the environment and any recent changes. We consider two components present in the line profiles of many sub-mm molecular lines, except for CO and SO which exhibit more complex line profiles. Radiative transfer modelling reveal that these two components have physically distinct properties, including column density, temperature, and line width. We associate the broad and narrow components with the redshifted and blueshifted lobes of outflowing gas, respectively, forming a bipolar structure. These lobes interact with pre-existing ambient material surrounding the central star through shocks, inducing non-uniform chemistry within the circumstellar environment. Via shocks, the redshifted lobe dissociates molecules such as CO to form HCO$^+$, whereas the blueshifted lobe excites H$_2$, which is detected via ro-vibrational lines. Shock models indicate that the blueshifted lobe shocks, where H$_2$ is excited, can be classified as J-type, and have a pre-shock density n$_H \geq$ 10$^7$ cm$^{-3}$. The detection of only HCO$^+$ in the redshifted lobe, and ro-vibrationally excited H$_2$ only in the blueshifted lobe, indicates different properties of the shocks, and therefore likely different kinematics in either the outflows themselves or in the properties of the ambient medium.

The diagnostics of the atomic emission suggest gas densities of 3 $\leq \log (N_e) \leq$ 5, and an electron temperature of 5\,000 $\leq$ $T_e$ $\leq$ 15\,000 K, which is unexpectedly high around a star with an expected effective temperature of around 3000 K. Modelling of the molecular bands present in the optical spectra reveal low rotational temperatures of 200--300 K. The moment-1 maps of CO and $^{29}$SiO support the presence of molecular outflows. Using sub-mm SO$_2$ emission as a thermometer, we derive the excitation temperature of the molecular gas to be 35--113 K. It likely represents the coolest emission regions of the remnant.

We therefore present V1309 Sco as a kinematically and chemically complex object. The presence of a bipolar outflow presents an analogy with other Galactic red novae such as V4332 Sgr and CK Vul \citep{kaminski2020ckvul,kaminski2021ckvul,mobeen2023}. The inconsistencies in distribution of the shock tracers and the kinematics of the bipolar lobes may indicate additional components to the circumstellar environment that we do not resolve, either spatially or spectroscopically. Further observations at higher angular resolution in sub-mm may reveal the puzzling nature of the outflows. Due to decreasing signal-to-noise in optical spectra, James Webb Space Telescope observations may serve as a better method to fully constrain the outflow and shock properties.

\begin{acknowledgements}
Based on observations collected at the European Southern Observatory under ESO programme 097.D-009 (PI T. Kami\'{n}ski). This paper makes use of the following ALMA data: ADS/JAO.ALMA2018.1.00336.S. ALMA is a partnership of ESO (representing its member states), NSF (USA) and NINS (Japan), together with NRC (Canada), MOST and ASIAA (Taiwan), and KASI (Republic of Korea), in cooperation with the Republic of Chile. The Joint ALMA Observatory is operated by ESO, AUI/NRAO and NAOJ.  T. S. and T. K. acknowledge funding from grant SONATA BIS no 2018/30/E/ST9/00398 from the Polish National Science Center (PI T. Kami\'{n}ski). We thank K. Menten, M.Z. Mobeen, R. Tylenda and the rest of the Mergestars team for their invaluable discussions.
\end{acknowledgements}
\bibliographystyle{aa} 
\bibliography{ref.bib}
\begin{appendix}
\onecolumn
\section{ALMA line table}\label{appendix: ALMA lines}
\begin{table}[h!]
    \centering
    \caption{Molecular transitions identified in the spectrum of V1309 Sco over 2 epochs.}
    \begin{tabular}{cccccc}\hline
    Molecule & Rest frequency (GHz) & E$_u$ (K) & Quantum numbers & Epoch 1? & Epoch 2?\\\hline
    CO & 345.796 & 33.19 & 3$\rightarrow$2 & $\checkmark$ & $\checkmark$ \\
    SO & 344.311 & 87.48 & 8(8)$\rightarrow$7(7) & $\checkmark$ & $\checkmark$ \\
    & 346.529 & 78.78 & 9(8)$\rightarrow$8(7) & $\checkmark$ & $\times$ \\
   $^{28}$SiO & 347.331 & 75.02 & 8$\rightarrow$7 & $\checkmark$ & $\times$\\
    $^{29}$SiO & 342.979 & 57.62 & 8$\rightarrow$7 & $\times$ & $\checkmark$ \\
    AlO & 344.454 & 82.77 & 9$\rightarrow$8 & $\checkmark$ & $\checkmark$ \\
    HCO$^+$ & 356.734 & 42.80 & 4$\rightarrow$3 & $\checkmark^a$ & $\checkmark$ \\
    AlOH & 346.153 & 99.69 & 11$\rightarrow$10 & $\checkmark$ & $\times$ \\\hline
    SO$_2$ & 345.339 & 92.98 & 13(2,12)$\rightarrow$12(1,11) & $\checkmark$ & $\checkmark$  \\
    & 345.449 & 521.00 & 26(9,17)$\rightarrow$27(8,20) & $\checkmark$ & $\checkmark$ \\
    & 346.523 & 164.47 & 16(4,12)$\rightarrow$16(3,13) & $\checkmark$ & $\checkmark$ \\
    & 346.653 & 168.14 & 19( 1,19)$\rightarrow$18( 0,18) & $\checkmark$ & $\checkmark$ \\
    & 355.046 & 111.00 & 12(4,8)$\rightarrow $12(3,9) & $\times$ & $\checkmark$ \\
    & 355.155 & 504.29 & 32(2,30)$\rightarrow$33(1,33) & $\checkmark$ & $\checkmark$ \\
    & 355.186 & 180.11 & 17(4,14)$\rightarrow$15(1,17) & $\checkmark$ & $\checkmark$ \\
    & 356.755 & 89.83 & 10(4,6)$\rightarrow$10(3,7) & $\checkmark$ & $\checkmark$ \\
    & 357.165 & 122.97 & 13(4,10)$\rightarrow$13(3,11) & $\checkmark$ & $\checkmark$ \\
    & 357.241 & 149.68 & 15(4,12)$\rightarrow$15(3,13) & $\checkmark$ & $\checkmark$ \\
    & 357.388 & 99.95 & 11(4,8)$\rightarrow$11(3,9) & $\checkmark$ & $\checkmark$ \\
    & 357.581 & 72.36 & 8(4,4)$\rightarrow$5(3,5) & $\checkmark$ & $\checkmark$ \\
    & 357.672 & 80.64 & 9(4,6)$\rightarrow$9(3,7) & $\checkmark$ & $\checkmark$ \\
    & 357.892 & 65.01 & 7(4,4)$\rightarrow$7(3,5) & $\checkmark$ & $\checkmark$ \\
    & 357.926 & 58.58 & 6(4,2)$\rightarrow$6(3,5) & $\checkmark$ & $\checkmark$ \\
    & 357.963 & 180.11 & 17(4,14)$\rightarrow$17(3,15) & $\checkmark$ & $\checkmark$ \\
    & 358.013 & 53.07 & 5(4,2)$\rightarrow$5(3,3) & $\checkmark$ & $\checkmark$ \\
    & 358.038 & 48.48 & 4(4,0)$\rightarrow$4(3,1) & $\checkmark$ & $\checkmark$ \\
    & 358.216 & 185.33 & 20(0,20)$\rightarrow$19(1,19) & $\checkmark$ & $\times$\\
    & 359.191 & 320.93 & 25(3,23)$\rightarrow$25(2,24) & $\checkmark$ & $\times$\\\hline
    \end{tabular}
    \tablefoot{$^a$ The HCO$^+$ line is covered by epoch 1 observations, but is not resolved from the SO$_2$ 10(4,6)$\rightarrow$10(3,7) line. Rest frequency, transition quantum numbers and upper level kinetic temperature (E$_u$) are specified, as taken from the JPL and CDMS catalogues accessible via CASSIS.}
    \label{table: ALMA lines}
\end{table}
\section{XSHOOTER atomic line table}\label{appendix: xshooter lines}
\begin{center}
\begin{longtable}{llllllll}
\caption{Identified atomic and H$_2$ emission lines in 2016 XSHOOTER spectrum.}\\
\hline $\lambda_{obs}$ & $\lambda_{em}^a$ & Identification &  Upper energy level & Integrated flux & Flux error & FWHM & Peak velocity\\
(\AA) & (\AA) & & (kK) & (erg/s/cm$^2$) & (erg/s/cm$^2$) & (km s$^{-1}$) & (km s$^{-1}$) \\\hline
\endfirsthead
\multicolumn{8}{c}{UVB}\\\hline
3960.09 & 3961.52 & Al I & 36.54 & 7.66$\times$10$^{-16}$ & 3.37$\times$10$^{-16}$ & 218.71 &--108.22 \\
4028.62 & 4030.75 & Mn I & 35.76 & 7.90$\times$10$^{-16}$ & 3.67$\times$10$^{-16}$ & 201.30 &--158.43\\
& 4033.06 & Mn I & 35.74 & 2.97$\times$10$^{-16}$ & 3.10$\times$10$^{-16}$ & 369.93 &\\
& 4034.48 & Mn I & 35.72 & 4.14$\times$10$^{-16}$ & 4.35$\times$10$^{-16}$ &  393.30 &\\
4066.56 & 4068.60 & [S II] &  35.42 & 7.13$\times$10$^{-16}$ & 2.96$\times$10$^{-16}$ & 166.32 &--150.32 \\
4214.26 & 4216.18 & Fe I &  34.18 & 1.14$\times$10$^{-15}$ & 3.51$\times$10$^{-16}$ &  250.84 &--136.53 \\
4224.91 & 4226.73 & Ca I & 34.10 & 2.09$\times$10$^{-15}$ & 4.26$\times$10$^{-16}$ & 324.22 &--129.09 \\
4252.55 & 4254.35 & Cr I & 33.88 & 1.78$\times$10$^{-15}$ & 2.79$\times$10$^{-16}$ &  230.25 &--126.84 \\
4273.27 & 4274.81 & Cr I &  33.72 & 1.31$\times$10$^{-15}$ & 2.91$\times$10$^{-16}$ &  229.62 &--108.00 \\
4287.99 & 4289.73 & Cr I &  33.60 & 1.69$\times$10$^{-15}$ & 2.90$\times$10$^{-16}$ &  252.47 &--121.60 \\
4374.24 & 4375.93 & Fe I &  32.94 &  1.98$\times$10$^{-15}$ & 2.49$\times$10$^{-16}$ & 197.59 &--115.78 \\
4569.10 & 4571.10 & Mg I &  31.53 & 6.77$\times$10$^{-16}$ & 2.50$\times$10$^{-16}$ &  169.35 &--131.17 \\
4605.14 & 4607.33 & Sr I & 31.28 & 1.10$\times$10$^{-15}$ & 1.96$\times$10$^{-16}$ &  184.22 &--142.50 \\
4859.27 & 4861.00 & H $\beta$ & 148.24 & 3.58$\times$10$^{-16}$ & 1.38$\times$10$^{-16}$ &  150.35 &--106.69 \\
5108.42 & 5110.41 & Fe I & 28.20 & 1.86$\times$10$^{-15}$ & 1.52$\times$10$^{-16}$ & 192.03 &--116.74 \\
5156.48 & 5158.78 & [Fe II] & 30.55 & 2.11$\times$10$^{-16}$ & 2.94$\times$10$^{-17}$ & 185.00 &--133.66 \\
5164.22 & 5166.28 & Fe I  & 27.90 & 1.36$\times$10$^{-15}$ & 1.59$\times$10$^{-16}$ &  205.51 &--119.54 \\
5202.48 & 5204.50 & Cr I &  38.64 & 1.86$\times$10$^{-15}$ & 1.26$\times$10$^{-16}$ &  208.38 &--116.36 \\
& 5206.02 & Cr I &  38.63 & 6.02$\times$10$^{-16}$ & 1.27$\times$10$^{-16}$ &  143.70 &  \\
& 5208.41 & Cr I &  38.62 & 8.82$\times$10$^{-17}$ & 9.93$\times$10$^{-17}$ &  189.11 &  \\
5258.77 & 5261.62 & [Fe II] & 30.81 &  1.18$\times$10$^{-16}$ & 7.64$\times$10$^{-17}$ & 144.92 &--162.39 \\
5261.53 & 5264.15 & Cr I & 38.64 & 3.58$\times$10$^{-16}$ & 9.53$\times$10$^{-17}$ & 143.70 &--149.21 \\
5294.63 & 5296.69 & Cr I & 38.64 & \multirow{2}{*}{3.86$\times$10$^{-16}$}
 & & &--116.77 \\
& 5298.27 & Cr I & 38.63  & & & \\
5344.07 & 5345.80 & Cr I  & 38.63 & 3.85$\times$10$^{-16}$ & 1.01$\times$10$^{-16}$ &  189.11 &--97.02 \\
5407.88 & 5409.78 & Cr I & 38.62 & 2.81$\times$10$^{-16}$ & 1.02$\times$10$^{-16}$ & 161.77 &--105.29 \\\hline
\multicolumn{8}{c}{VIS}\\\hline
5860.20 & & U\footnote{U indicate unidentified lines.} &   & 3.82$\times$10$^{-16}$ & 1.36$\times$10$^{-16}$ &  132.30 &  \\
5886.44 & 5889.95 & Na I  & 24.47 &5.30$\times$10$^{-15}$ & 1.42$\times$10$^{-16}$ &   173.42 &--178.66 \\
5892.77 & 5895.92 & Na I  & 24.45 &5.02$\times$10$^{-15}$ & 1.22$\times$10$^{-16}$ &   152.98 &--160.17 \\
6297.13 & 6300.30 & [O I]  & 22.88 &6.80$\times$10$^{-15}$ & 6.73$\times$10$^{-17}$ &   105.74 &--150.84 \\
6360.30 & 6363.78 & [O I]  & 22.88 & 2.05$\times$10$^{-15}$ & 6.64$\times$10$^{-17}$ & 112.22 &--163.94 \\
6559.94 & 6562.80 & H$\alpha$  & 140.55 &3.04$\times$10$^{-15}$ & 6.39$\times$10$^{-17}$ &   126.91 &--130.65 \\
6569.95 & 6572.78 & [Ca I & 21.93 &3.41$\times$10$^{-15}$ & 7.25$\times$10$^{-17}$ &   162.40 &--129.08 \\
6704.65 & 6707.81 & Li I &  21.49 &4.78$\times$10$^{-16}$ & 1.19$\times$10$^{-16}$ &   129.27 &--141.23 \\
6712.49 & 6716.44 & [S II] & 21.46 & 2.58$\times$10$^{-16}$ & 6.61$\times$10$^{-17}$ & 127.39 &--176.31 \\
6727.13 & 6730.82 & [S II]  & 21.41 &4.41$\times$10$^{-16}$ & 6.38$\times$10$^{-17}$ &   118.35 &--164.35 \\
7151.27 & 7155.17 & [Fe II] &  22.84 &4.29$\times$10$^{-16}$ & 4.55$\times$10$^{-17}$ &   99.94 &--163.41 \\
7168.29 & 7172.00 & [Fe II] &   23.60 &2.58$\times$10$^{-16}$ & 6.16$\times$10$^{-17}$ &   120.12 &--155.08 \\
7287.36 & 7291.47 & [Ca II] &   19.77 &3.14$\times$10$^{-16}$ & 7.84$\times$10$^{-17}$ &   168.71 &--168.98 \\
7320.01 & 7323.89 & [Ca II] &   19.68 &5.72$\times$10$^{-16}$ & 5.63$\times$10$^{-17}$ &   190.65 &--158.82 \\
7586.66 &  & U  &  & 1.59$\times$10$^{-16}$ & 4.12$\times$10$^{-17}$ & 129.81 &  \\
7661.93 & 7664.90 & K I & 18.80 & 4.45$\times$10$^{-14}$ & 1.31$\times$10$^{-16}$ &  182.34 &--116.16 \\
7695.74 & 7698.96 & K I &  18.72 & 5.09$\times$10$^{-14}$ & 1.38$\times$10$^{-16}$ &  202.57 &--125.38 \\
7797.09 & 7800.27 & Rb I &  18.48 & 5.09$\times$10$^{-15}$ & 4.66$\times$10$^{-17}$ &   169.99 &--122.22 \\
7944.52 & 7947.60 & Rb I & 18.13 & 4.08$\times$10$^{-15}$ & 4.53$\times$10$^{-17}$ &   178.72 &--116.18 \\
8056.77 &  & U &   & 4.70$\times$10$^{-16}$ & 5.49$\times$10$^{-17}$ &  230.63 & \\
8601.38 & & U &   & 3.48$\times$10$^{-16}$ & 3.45$\times$10$^{-17}$ &  102.02 & \\
8612.21 & 8616.95 & [Fe II] & 19.43 & 4.26$\times$10$^{-16}$ & 3.05$\times$10$^{-17}$ &  127.72 &--164.91 \\
8662.33 &  & U & & 5.65$\times$10$^{-17}$ & 2.89$\times$10$^{-17}$ &  56.90 & \\ 
8722.00 & & U & &  2.62$\times$10$^{-16}$ & 3.08$\times$10$^{-17}$ &  88.79 & \\
9220.98 & & U &  & 3.22$\times$10$^{-16}$ & 3.98$\times$10$^{-17}$ & 200.28 & \\\hline
\multicolumn{8}{c}{NIR} \\\hline
10281.40 & 10286.73 & [S II] & 35.42 & 3.63$\times$10$^{-16}$ & 3.16$\times$10$^{-17}$ &  108.06 &--155.34\\
10291.90 &  & U &   & 6.69$\times$10$^{-16}$ & 3.21$\times$10$^{-17}$ &  112.30 & \\
10295.80 &  & U &   & 1.49$\times$10$^{-15}$ & 2.30$\times$10$^{-17}$ &  57.45 & \\
10300.10 &  & U &   & 3.35$\times$10$^{-16}$ & 2.92$\times$10$^{-17}$ &  82.46 & \\
10314.90 & 10320.49 & [S II] & 35.42 & 4.76$\times$10$^{-16}$ & 2.04$\times$10$^{-17}$ &  116.90 &--162.38 \\
10331.10 & 10336.41 & [S II] & 35.36 & 2.82$\times$10$^{-16}$ & 2.08$\times$10$^{-17}$ & 114.63 &--154.01 \\
10365.00 & 10370.49 & [S II] & 35.36 & 1.06$\times$10$^{-16}$ & 1.72$\times$10$^{-17}$ &  75.26 &--158.71 \\
10392.10 & & U &  & 4.72$\times$10$^{-16}$ & 1.86$\times$10$^{-17}$ &  101.43 & \\
10401.8 & & U &  & 3.41$\times$10$^{-16}$ & 1.92$\times$10$^{-17}$ &  98.48 & \\
10926.9 & & U &   & 1.17$\times$10$^{-15}$ & 1.58$\times$10$^{-17}$ &  98.55 & \\
21202.70 & 21218.00 & H$_2$ 1--0 S(1) & & 2.66$\times$10$^{-16}$ & 1.93$\times$10$^{-17}$ & 105.52 & -216.18 \\
24048.70 & 24066.00 & H$_2$ 1--0 Q(1) & & 5.11$\times$10$^{-15}$ & 1.04$\times$10$^{-16}$ &  123.88 &--215.51 \\\hline
\label{table: xshooter lines}
\end{longtable}
\end{center}
\newpage\twocolumn
\section{Molecule identification and modelling}\label{appendix: xshooter molecules}
Identification of molecular bands relied on comparison to previous studies \citep{kaminski2015} as well as using the interactive molecular modelling tool Pgopher \footnote{http://pgopher.chm.bris.ac.uk/} \citep{western2017pgopher} to model the detected molecules. Although the identification of molecular lines and bands are more difficult than for atomic lines, the specific shape of some molecular bands help to identify them. 

We generated models from the molecular line lists available in 2022 on the ExoMol online database\footnote{https://www.exomol.com/data/} \citep{patrascu2015AlO,mckemmish2016VO,mckemmish2019Tio,chubb2021TiO,chubb2021VO}. The models are calculated using both the vibrational (T$_{vib}$) and rotational temperature (T$_{rot}$) as well as the quantum properties for each ro-vibrational state. ExoMol files are available for AlO, TiO, and VO. We also modelled ScO and CrO using custom-made files generated from published spectroscopic constants of the electronic bands covered (see \cite{kaminski2009} for details). For each molecule, a model grid was generated between 100 and 3500 K for both T$_{rot}$ and T$_{vib}$ at intervals of 100 K, with the condition that T$_{rot} <$ T$_{vib}$. As there are regions of the spectrum where certain molecules are not observed, the models were not generated across the entire spectrum for all molecules to save time. AlO was modelled from 5000 to 25000 \AA, TiO was modelled from 4880 to 8700 \AA\space and VO is modelled between 3400 and 16000 \AA. The simulations were then fitted via minimised residuals to find the best fitting rotational and vibrational temperatures.

\begin{figure*}[h!]
    \centering
    \includegraphics[width=1.1\textwidth, trim={2.5cm 0cm 0 2.5cm}, clip]{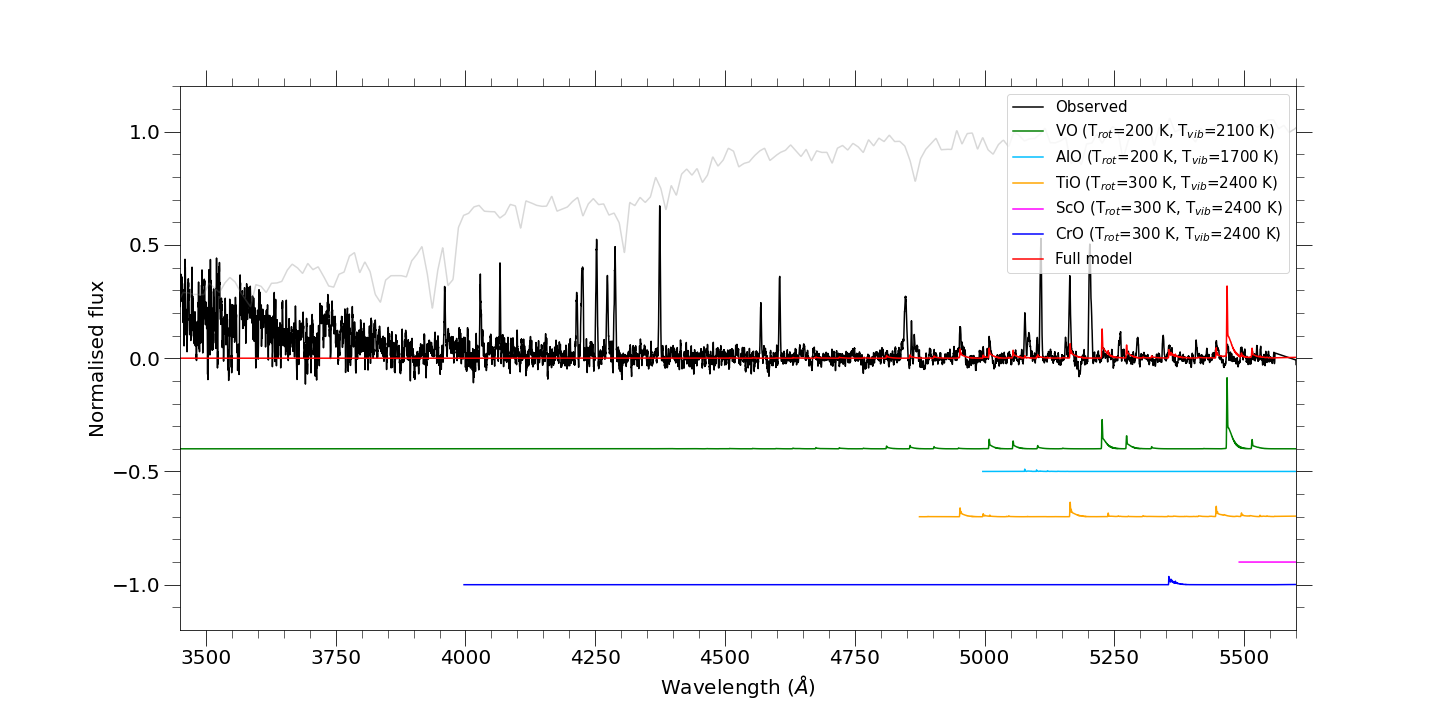}
    \includegraphics[width=1.1\textwidth, trim={2.5cm 0cm 0 2.5cm}, clip]{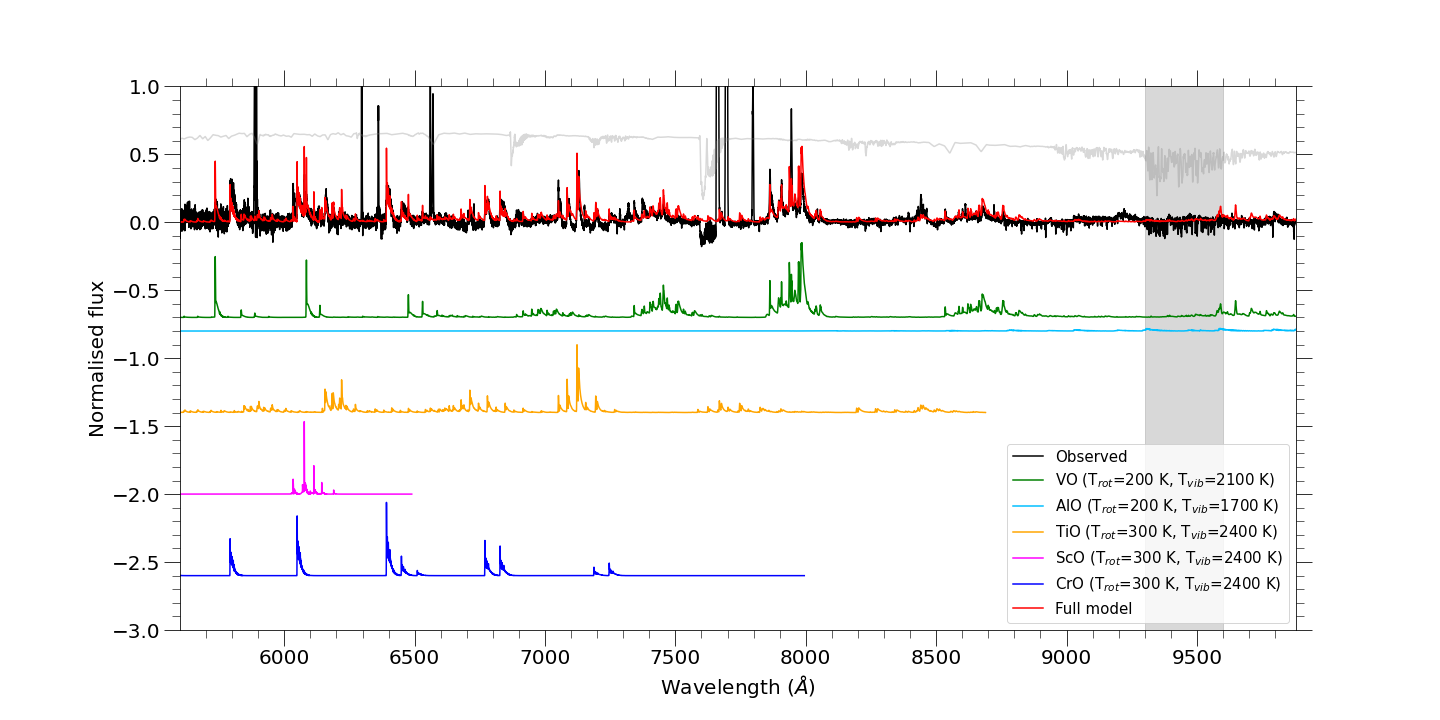}
    \caption{Plot of the XSHOOTER UVB (\textbf{top}) and VIS (\textbf{bottom}) spectra and best-fitting molecular model. The red line shows the full model with all components added, and the green, light blue, gold, magenta and dark blue lines underneath represent the best-fitting VO, AlO, TiO, ScO and CrO models respectively. The atmospheric transmission spectrum is represented by the grey curve and the grey regions represent the strongest telluric absorption bands.}
    \label{fig: molec models page 1}
\end{figure*}

\begin{figure}[h!]
    \centering
    \includegraphics[width=1.1\textwidth, trim={2.5cm 0cm 0 2.5cm}, clip]{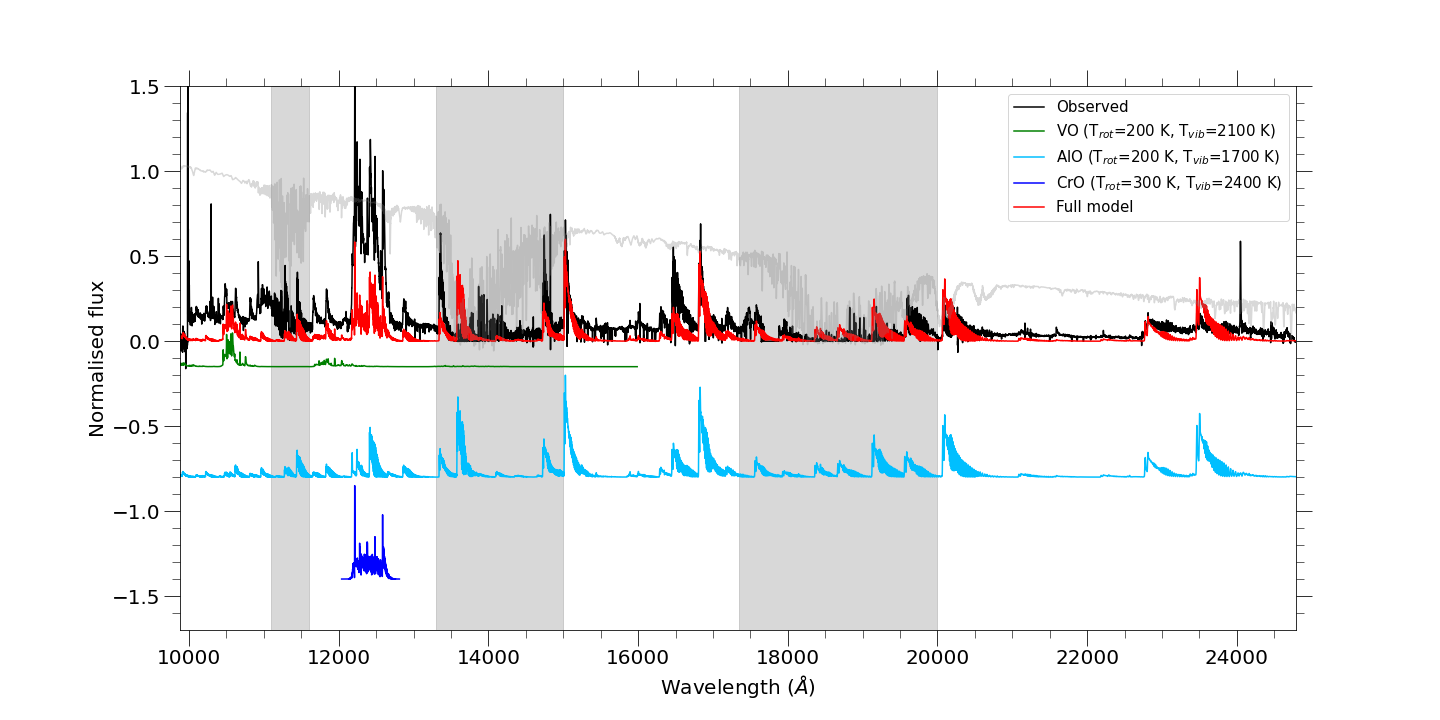}
    \caption{Same as Fig. \ref{fig: molec models page 1} but for the XSHOOTER NIR spectrum.}
    \label{fig: molec models page 2}
\end{figure}

The fitting routine was carried out across prominent features of each molecule. The fitting region for AlO from 14200 \AA\space to 17400 \AA\space covered multiple AlO A$^2\Pi$--X$^2\Sigma^+$ bands. For TiO, the fitting region we used was from 6110 \AA\space to 7260 \AA, covering multiple TiO $\gamma$ features including the prominent TiO $\gamma$ (0,0) band between 7040 and 7160 \AA. For VO, we used the region between 7090 and 9010 \AA\space to fit the VO models to the spectrum. This covered multiple VO B$^4\Pi$--X$^4\Sigma^-$ bands in the VIS spectrum. The CrO B$^5\Pi$--X$^5\Pi$ band was modelled across 6380 and 6500 \AA. Although this range is narrow, it is the only CrO B$^5\Pi$--X$^5\Pi$ emission that is uncontaminated from atoms or other molecules. CrO A$^5\Sigma^{+}$--X$^5\Pi$ bands were modelled across the main CrO NIR feature between 12185 and 12655 \AA. Finally, ScO was modelled across the only system detected, A$^2\Pi$--X$^2\Sigma^{+}$, prominent between 5610 and 6340 \AA. The full ExoMol model, together with the individual components of each molecule, is presented in Figs. \ref{fig: molec models page 1}--\ref{fig: molec models page 2}.

Due to issues with the fitting of VO, CrO, and ScO, we are unable to present reliable constraints on the rotational and vibrational temperatures of these molecules. The output of the fitting routine returned the maximum possible T$_{vib}$ of 3500 K that we simulated in our model grid, no matter what input parameters were used. We extended the model grids to higher temperatures, but only achieved convergence at unrealistic temperatures of $\sim$5000 K. Furthermore, upon visual examination, the fit residuals were larger than those at T$_{vib}\leq$ 3500 K for many prominent features. Therefore, we instead present in Figs. \ref{fig: molec models page 1}--\ref{fig: molec models page 2} the VO model simulated with the average ro-vibrational temperatures that we found between TiO and AlO.

The identification of features of the AlO spectrum presented in this paper is improved compared to that presented in \cite{kaminski2015} owing to the more complete list of lines in the ExoMol database. In particular, a feature near 12200 \AA\ marked in \cite{kaminski2015} as an AlO perturbation turned out to be a ro-vibrational $\varv$=9--0 band of AlO X$^{2}\Sigma^{+}$  (within the ground electronic state of AlO).
\onecolumn
\newpage
\begin{center}
\begin{longtable}{llll}
\caption{Molecular bands observed in V1309 Sco in 2016 using XSHOOTER.}\\
\hline
$\lambda_{start}$ (\AA) & $\lambda_{end}$ (\AA) & $\lambda_{obs}$ (\AA) & Molecular state\\\hline
\multicolumn{4}{c}{UVB}\\\hline
4841.00 & 4855.00 & 4847.50 & TiO $\alpha$ (4,2) \\
4862.85 & 4868.50 & 4864.40 & AlO B$^2\Sigma^+$--X$^2\Sigma^+$ (1,1) \\
4948.50 & 4968.50 & 4952.90 & TiO $\alpha$ (1,0) \\
5007.00 & 5017.65 & 5009.00 & VO C$^4\Sigma^-$--X$^4\Sigma^-$ (3,0) \\
\multirow{2}{*}{5075.00} & \multirow{2}{*}{5090.20} & 5077.80 & \multirow{2}{*}{AlO B$^2\Sigma^+$--X$^2\Sigma^+$ (0,1)} \\
& & 5085.80 & \\
5098.00 & 5105.00 & 5101.38 & AlO B$^2\Sigma^+$--X$^2\Sigma^+$ (1,2)\\
5349.80 & 5370.30 & 5355.00 & CrO B$^5\Pi$--X$^5\Pi$ (3,0) \\
5442.50 & 5458.75 & 5446.35 & TiO $\alpha$ (0,1) \\\hline
\multicolumn{4}{c}{VIS}\\\hline
5462.30 & 5483.00 & 5467.30 & VO C$^4\Sigma^-$--X$^4\Sigma^-$ (1,0)\\
5489.90 & 5509.00 & 5494.40 & TiO $\alpha$ (1,2)\\
5731.25 & 5746.10 & 5734.70 & VO C$^4\Sigma^-$--X$^4\Sigma^-$ (0,0)\\
5789.20 & 5826.30 & 5793.64 & CrO B$^5\Pi$--X$^5\Pi$ (1,0)\\
5841.25 & 5854.60 & 5846.60 & TiO $\gamma^{\prime}$ (1,0)\\
6030.00 & 6045.70 & 6033.90 & ScO A$^2\Pi_{3/2}$--X$^2\Sigma^+$ (0,0)\\
6046.70 & 6080.00 & 6049.50 & CrO B$^5\Pi_{-1}$--X$^5\Pi_{-1}$ (0,0)\\
\multirow{2}{*}{6068.20} & \multirow{2}{*}{6088.88} & \multirow{2}{*}{6076.63} & ScO A$^2\Pi_{1/2}$--X$^2\Sigma^+$ (0,0) (1,1)\\
& & & VO C$^4\Sigma^-$--X$^4\Sigma^-$\\
6150.25 & 6232.50 & 6189.30 & TiO $\gamma^{\prime}$ (0,0)\\
\multirow{3}{*}{6387.80} & \multirow{3}{*}{6445.15} & 6392.50 & CrO B$^5\Pi_{-1}$--X$^5\Pi_{-1}$ (0,1)\\
& & 6398.50 & CrO B$^5\Pi_{1}$--X$^5\Pi_{1}$ (0,1)\\
& & 6404.00 & CrO B$^5\Pi_{2}$--X$^5\Pi_{2}$ (0,1)\\
6446.50 & 6495.00 & 6474.70 & CrO B$^5\Pi$--X$^5\Pi$ (1,2)\\
6611.40 & 6636.56 & 6618.00 & TiO $\gamma^{\prime}$ (1,2) F$_1$--F$_1$\\
\multirow{2}{*}{6645.60} & \multirow{2}{*}{6661.69} & \multirow{2}{*}{6647.20} & TiO $\gamma$ (1,0) F$_3$--F$_3$\\
& & & + TiO $\gamma^{\prime}$ (1,2) F$_2$--F$_2$\\
\multirow{2}{*}{6671.55} & \multirow{2}{*}{6700.28} & \multirow{2}{*}{6678.80} & TiO $\gamma$ (1,0) F$_2$--F$_2$\\
& & & + TiO $\gamma^{\prime}$ (1,2) F$_3$--F$_3$\\
\multirow{2}{*}{6708.42} & \multirow{2}{*}{6724.00} & \multirow{2}{*}{6712.49} & TiO $\gamma$ (1,0) F$_1$--F$_1$\\
& & & + TiO $\gamma^{\prime}$ (2,1) F$_3$--F$_3$\\
\multirow{3}{*}{6765.30} & \multirow{3}{*}{6800.00} & \multirow{3}{*}{6779.30} & CrO B$^5\Pi_{-1}$--X$^5\Pi_{-1}$(0,2)\\
& & & TiO $\gamma$ (2,1) F$_1$--F$_1$\\
& & & CrO B$^5\Pi_3$--X$^5\Pi_3$ (0,2)\\
\multirow{2}{*}{6822.40} & \multirow{2}{*}{6874.75} & \multirow{2}{*}{6829.10} & CrO B$^5\Pi$--X$^5\Pi$ (1,3)\\
& & & + TiO b$^1\Pi$--X$^3\Pi$ (0,0)\\
6877.85 & 6948.60 & 6891.10 & CrO B$^5\Pi$--X$^5\Pi$ (2,0)\\
7047.75 & 7068.15 & 7058.50 & TiO $\gamma$ (0,0) F$_3$--F$_3$\\
7080.00 & 7107.75 & 7091.21 & TiO $\gamma$ (0,0) F$_2$--F$_2$\\
7116.20 & 7149.25 & 7130.01 & TiO $\gamma$ (0,0) F$_1$--F$_1$\\
7154.50 & 7166.30 & 7156.80 & TiO $\gamma$ (1,1) F$_2$--F$_2$\\
7192.10 & 7221.20 & 7201.30 & TiO $\gamma$ (1,1) + TiO $\gamma$ (2,2)\\
7241.65 & 7284.50 & 7248.00 & CrO B$^5\Pi$--X$^5\Pi$ (1,4)\\
\multirow{4}{*}{7329.30} & \multirow{4}{*}{7486.10} & 7342.50 & VO B$^4\Pi_{5/2}$--X$^4\Sigma^-$ (1,0)\\
& & 7381.10 & VO B$^4\Pi_{3/2}$--X$^4\Sigma^-$ (1,0)\\
& & 7407.00 & VO B$^4\Pi_{1/2}$--X$^4\Sigma^-$ (1,0)\\
& & 7454.72 & VO B$^4\Pi_{-1/2}$--X$^4\Sigma^-$ (1,0)\\
\multirow{4}{*}{7840.50} & \multirow{4}{*}{8025.25} & 7862.50 & VO B$^4\Pi_{5/2}$--X$^4\Sigma^-$ (0,0)\\
& & 7907.28 & VO B$^4\Pi_{3/2}$--X$^4\Sigma^-$ (0,0)\\
& & 7937.59 & VO B$^4\Pi_{1/2}$--X$^4\Sigma^-$ (0,0)\\
& & 7984.29 & VO B$^4\Pi_{-1/2}$--X$^4\Sigma^-$ (0,0)\\
\multirow{3}{*}{8350.00} & \multirow{3}{*}{8481.30} & 8394.00 & CrO A$^{\prime 5}\Delta_2$--X$^5\Pi_1$ (0,0)\\
& & 8425.81 & CrO A$^{\prime 5}\Delta_1$--X$^5\Pi_0$ (0,0)\\
& & 8463.30 & CrO A$^{\prime 5}\Delta_0$--X$^5\Pi_{-1}$ (0,0)\\
8430.80 & 8458.20 & 8441.42 & TiO $\epsilon$ (0,0)\\
\multirow{4}{*}{8515.00} & \multirow{4}{*}{8710.00} & 8534.90 & VO B$^4\Pi_{5/2}$--X$^4\Sigma^-$ (0,1)\\
& & 8586.50 & VO B$^4\Pi_{3/2}$--X$^4\Sigma^-$ (0,1)\\
& & 8633.50 & VO B$^4\Pi_{1/2}$--X$^4\Sigma^-$ (0,1)\\
& & 8679.30 & VO B$^4\Pi_{-1/2}$--X$^4\Sigma^-$ (0,1)\\
9908.15 & 9980.00 & 9918.50 & AlO A$^2\Pi_{3/2}$--X$^2\Sigma^+$ (7,0)\\\hline
\multicolumn{4}{c}{NIR}\\\hline
10208.30 & 10276.90 & 10229.00 & AlO A$^2\Pi_{3/2}$--X$^2\Sigma^+$ (8,1)\\
\multirow{4}{*}{10451.10} & \multirow{4}{*}{10653.30} & 10457.10 & VO A$^4\Pi_{5/2}$--X$^4\Sigma^-$ (0,0)\\
& & 10502.10 & VO A$^4\Pi_{3/2}$--X$^4\Sigma^-$ (0,0)\\
& & 10559.40 & VO A$^4\Pi_{1/2}$--X$^4\Sigma^-$ (0,0)\\
& & 10579.70 & VO A$^4\Pi_{-1/2}$--X$^4\Sigma^-$ (0,0)\\
\multirow{2}{*}{10464.70} & \multirow{2}{*}{10705.10} & 10483.00 & AlO A$^2\Pi_{1/2}$--X$^2\Sigma^+$ (6,0)\\
& & 10623.40 & AlO A$^2\Pi_{3/2}$--X$^2\Sigma^+$ (6,0)\\
\multirow{2}{*}{10802.70} & \multirow{2}{*}{11000.00} & 10815.70 & AlO A$^2\Pi_{1/2}$--X$^2\Sigma^+$ (7,1)\\
& & 10976.40 & AlO A$^2\Pi_{3/2}$--X$^2\Sigma^+$ (7,1)\\
\multirow{2}{*}{11270.40} & \multirow{2}{*}{11645.50} & 11282.20 & AlO A$^2\Pi_{1/2}$--X$^2\Sigma^+$ (5,0)\\
& & 11446.50 & AlO A$^2\Pi_{3/2}$--X$^2\Sigma^+$ (5,0)\\
\multirow{2}{*}{11645.50} & \multirow{2}{*}{12032.90} & 11664.00 & AlO A$^2\Pi_{1/2}$--X$^2\Sigma^+$ (6,1)\\
& & 11838.20 & AlO A$^2\Pi_{3/2}$--X$^2\Sigma^+$ (6,1)\\
12170.60 & 12186.70 & 12180.50 & AlO X$^2\Sigma^+$--X$^2\Sigma^+$ (9,0)\\
\multirow{2}{*}{12236.10} & \multirow{2}{*}{12737.00} & 12246.00 & AlO A$^2\Pi_{1/2}$--X$^2\Sigma^+$ (4,0)\\
& & 12422.20 & AlO A$^2\Pi_{3/2}$--X$^2\Sigma^+$ (4,0)\\
\multirow{5}{*}{12186.50} & \multirow{5}{*}{12655.50} & 12215.10 & CrO A$^5\Sigma^+$--X$^5\Pi_{-1}$ (0,0)\\
& & 12285.70 & CrO A$^5\Sigma^+$--X$^5\Pi_{0}$ (0,0)\\
& & 12382.80 & CrO A$^5\Sigma^+$--X$^5\Pi_{1}$ (0,0)\\
& & 12485.20 & CrO A$^5\Sigma^+$--X$^5\Pi_{2}$ (0,0)\\
& & 12588.30 & CrO A$^5\Sigma^+$--X$^5\Pi_{3}$ (0,0)\\
12854.30 & 12916.30 & 12871.10 & AlO A$^2\Pi_{3/2}$--X$^2\Sigma^+$ (5,1)\\
13337.00 & 13469.90 & 13356.90 & AlO A$^2\Pi_{1/2}$--X$^2\Sigma^+$ (3,0)\\
\multirow{2}{*}{14717.90} & \multirow{2}{*}{15244.60} & 14744.80 & AlO A$^2\Pi_{1/2}$--X$^2\Sigma^+$ (2,0)\\
& & 15033.90 & AlO A$^2\Pi_{3/2}$--X$^2\Sigma^+$ (2,0)\\
15434.20 & 15465.80 & 15441.30 & AlO A$^2\Pi_{1/2}$--X$^2\Sigma^+$ (7,4)\\
\multirow{2}{*}{15982.00} & \multirow{2}{*}{16442.00} & 16005.40 & AlO A$^2\Pi_{1/2}$--X$^2\Sigma^+$ (4,2)\\
& & 16302.70 & AlO A$^2\Pi_{3/2}$--X$^2\Sigma^+$ (4,2)\\
\multirow{2}{*}{16445.50} & \multirow{2}{*}{17001.40} & 16471.20 & AlO A$^2\Pi_{1/2}$--X$^2\Sigma^+$ (1,0)\\
& & 16825.90 & AlO A$^2\Pi_{3/2}$--X$^2\Sigma^+$ (1,0)\\
17161.80 & 17708.40 & 10195.70 & AlO $A^2\Pi_{1/2}$--X$^2\Sigma^+$ (2,1)\\
\multirow{2}{*}{19550.60} & \multirow{2}{*}{20556.00} & 19586.90 & AlO A$^2\Pi_{1/2}$--X$^2\Sigma^+$ (1,1)\\
& & 20096.00 & AlO A$^2\Pi_{3/2}$--X$^2\Sigma^+$ (1,1)\\
\multirow{2}{*}{21575.30} & \multirow{2}{*}{21656.20} & \multirow{2}{*}{21587.70} & either AlO A$^2\Pi_{1/2}$--X$^2\Sigma^+$ (7,6)\\
& & & or AlO A$^2\Pi_{1/2}$--X$^2\Sigma^+$ (3,3)\\
\multirow{2}{*}{22771.00} & \multirow{2}{*}{23432.00} & 22774.50 & AlO A$^2\Pi_{1/2}$--X$^2\Sigma^+$ (0,1)\\
& & 22811.40 & + AlO A$^2\Pi_{1/2}$--X$^2\Sigma^+$ (4,4)\\
\multirow{2}{*}{23451.50} & \multirow{2}{*}{23694.00} & 23462.10 & AlO A$^2\Pi_{3/2}$--X$^2\Sigma^+$ (0,1)\\
& & 23502.80 & + AlO A$^2\Pi_{3/2}$--X$^2\Sigma^+$ (4,4)\\\hline
\label{table: molecular bands}
\end{longtable}
\end{center}
\newpage
\section{Molecular models}\label{appendix: molecular models}
\newpage
\section{Epoch comparison}\label{appendix: epoch comparison}
\begin{figure}[h!]
    \centering
    \includegraphics[width=1.1\textwidth, trim={2.5cm 0cm 0 2.5cm}, clip]{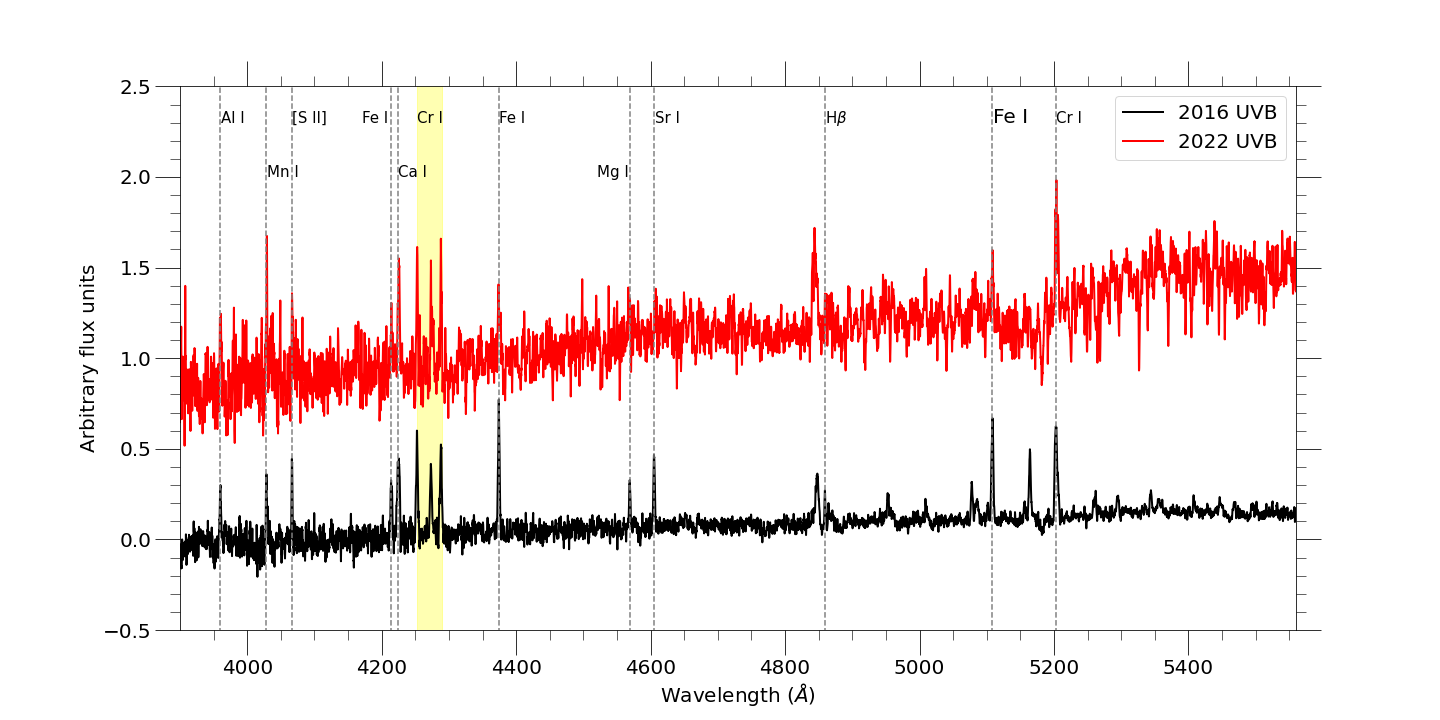}
    \caption{XSHOOTER UVB spectra of V1309 Sco taken in 2016 (black), and 2022 (red). Prominent spectral lines are indicated by vertical grey lines. The Cr I triplet at $\sim$4270 \AA\space is highlighted by the yellow area in the top panel. The 2022 spectrum is shifted along the y-axis.}
    \label{fig: xshooter epoch comparison 1}
\end{figure}
\newpage
\begin{figure}[h!]
    \centering
    \includegraphics[width=1.1\textwidth, trim={2.5cm 0cm 0 2.5cm}, clip]{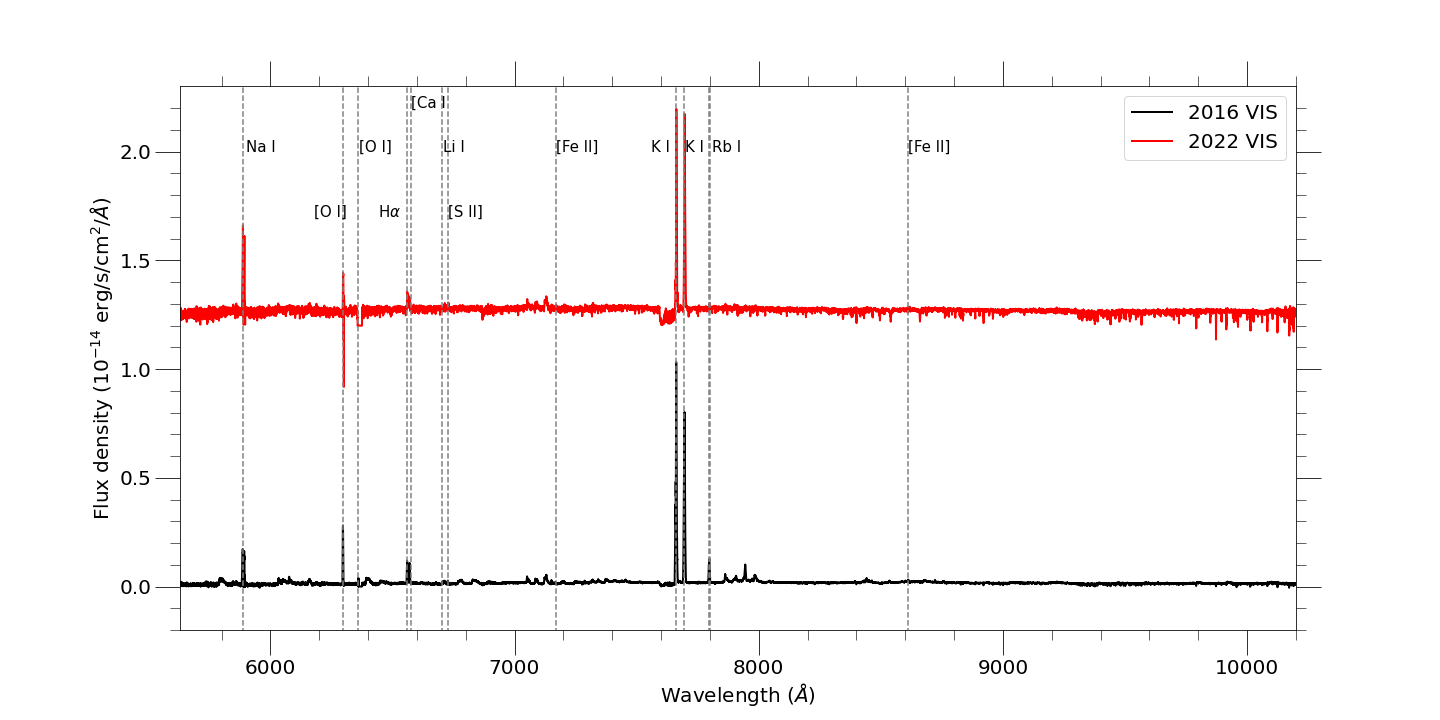}
    \includegraphics[width=1.1\textwidth, trim={2.5cm 0cm 0 2.5cm}, clip]{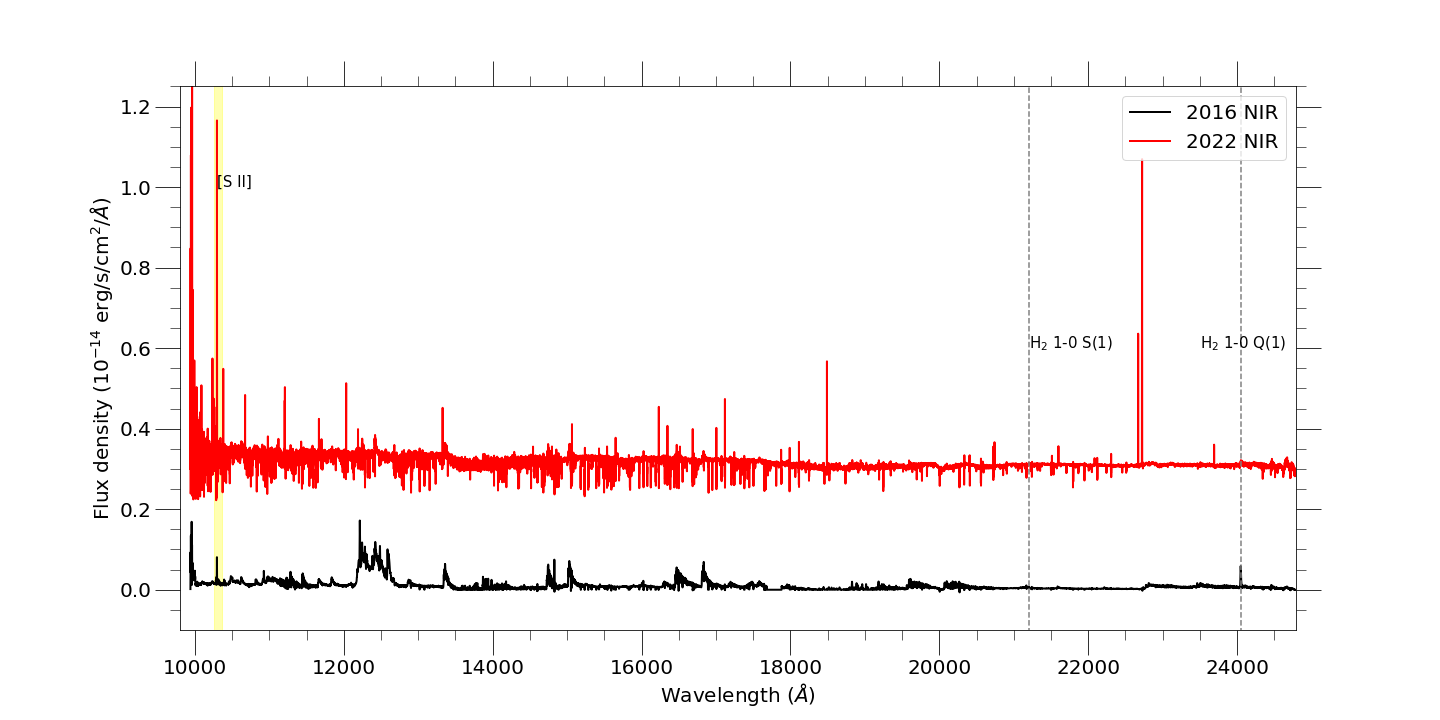}
    \caption{Same as Fig. \ref{fig: xshooter epoch comparison 1}, but for the VIS and NIR spectra in the top and bottom panels, respectively. The yellow area in the bottom panel highlights the four [S II] lines found between 10280--10370\AA.}
    \label{fig: xshooter epoch comparison 2}
\end{figure}
\twocolumn
\section{H$_2$ models}\label{appendix: H2 models}
\begin{figure}[h!]
    \centering
    \includegraphics[width=\columnwidth, trim={3.5cm 1cm 3.5cm 1cm, clip}]{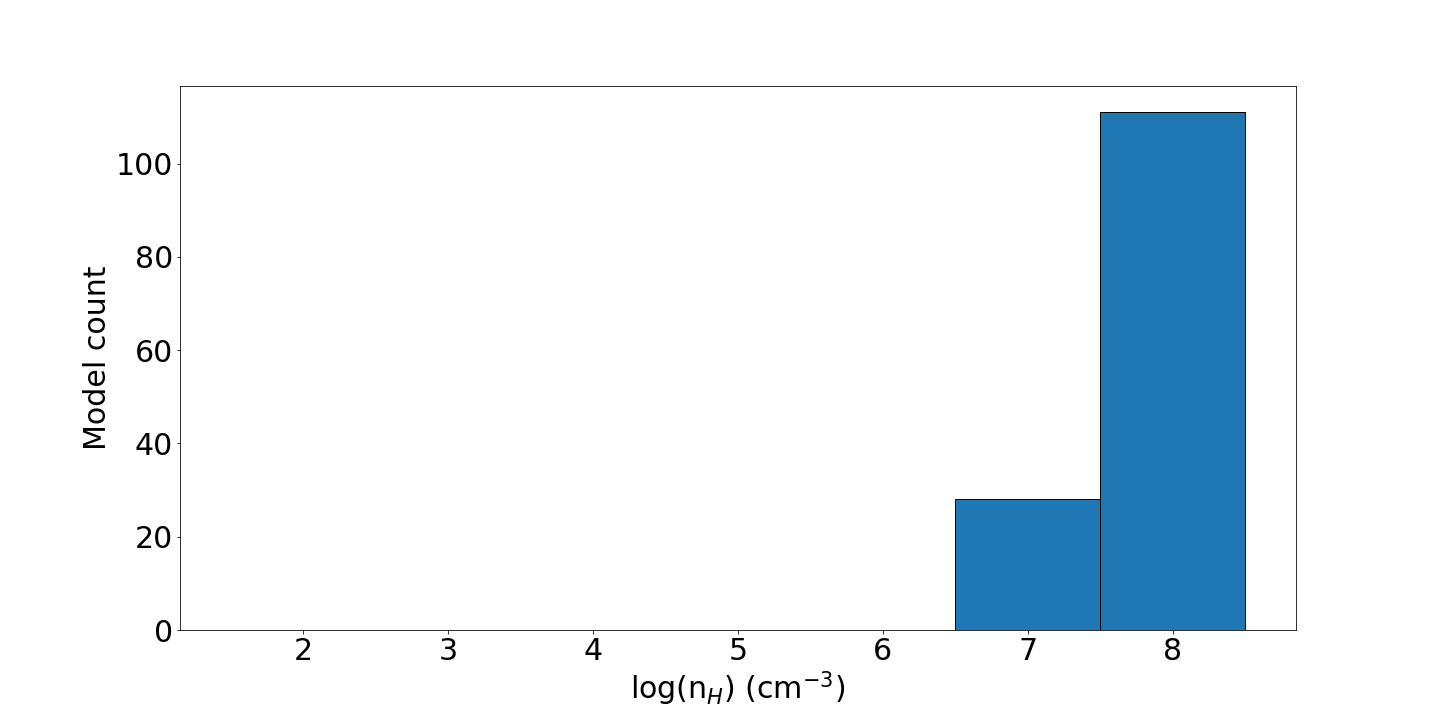}
    \includegraphics[width=\columnwidth, trim={3.5cm 1cm 3.5cm 1cm, clip}]{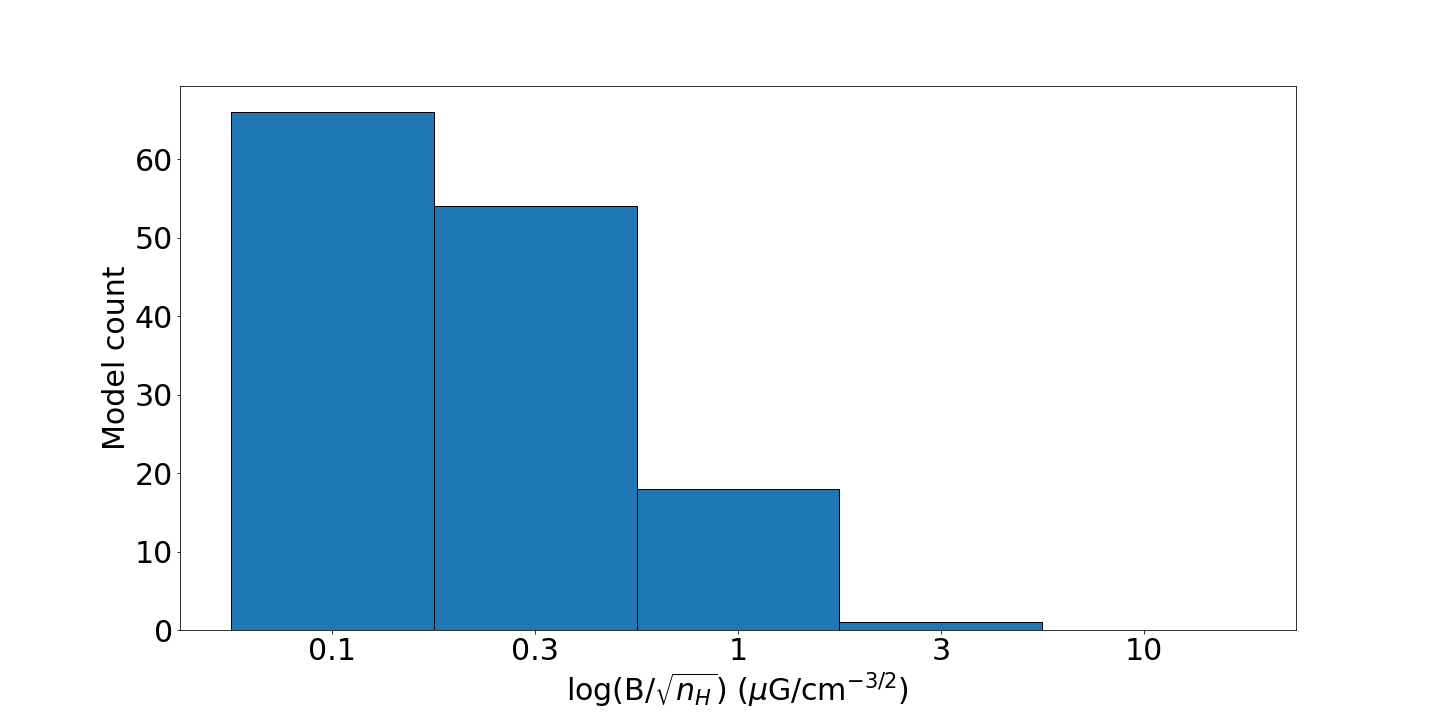}
\end{figure}

\begin{figure}[h!]
    \centering
    \includegraphics[width=\columnwidth, trim={3.5cm 1cm 3.5cm 1cm, clip}]{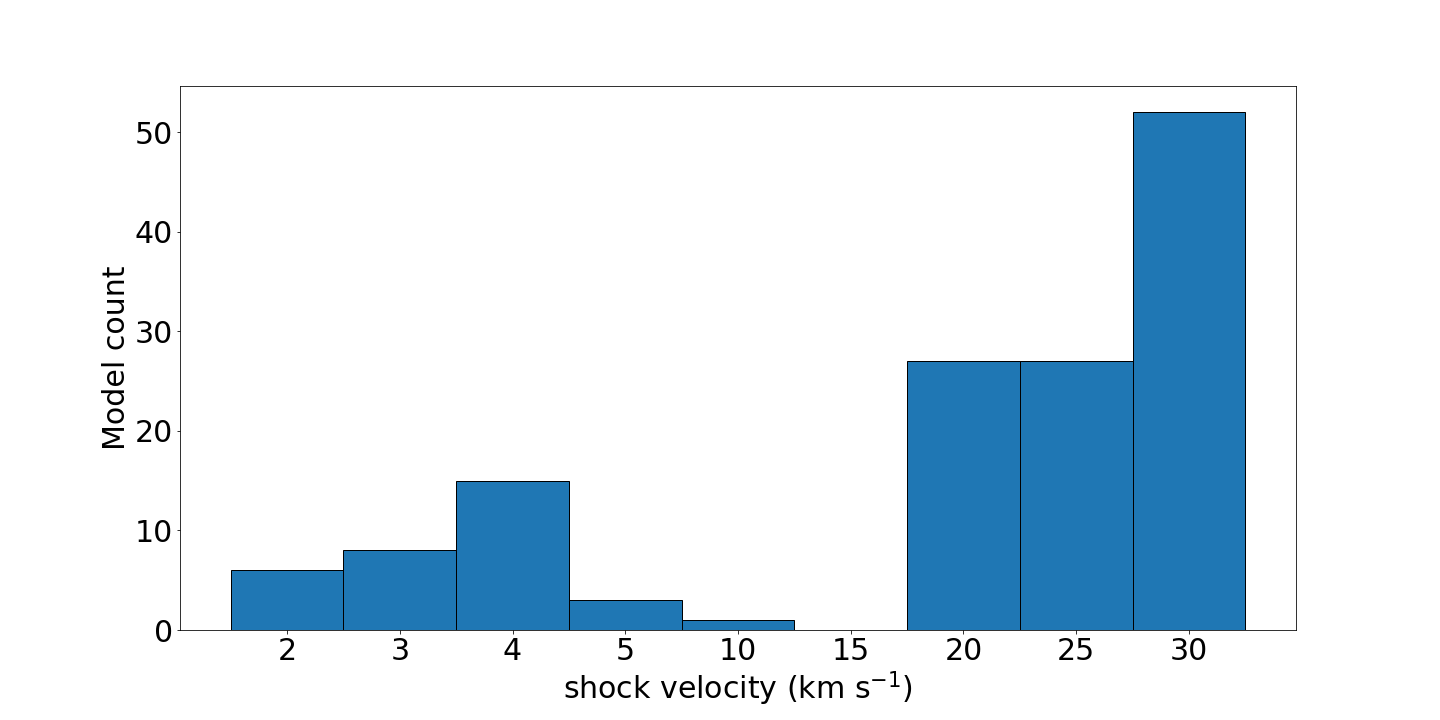}
    \includegraphics[width=\columnwidth, trim={3.5cm 1cm 3.5cm 1cm, clip}]{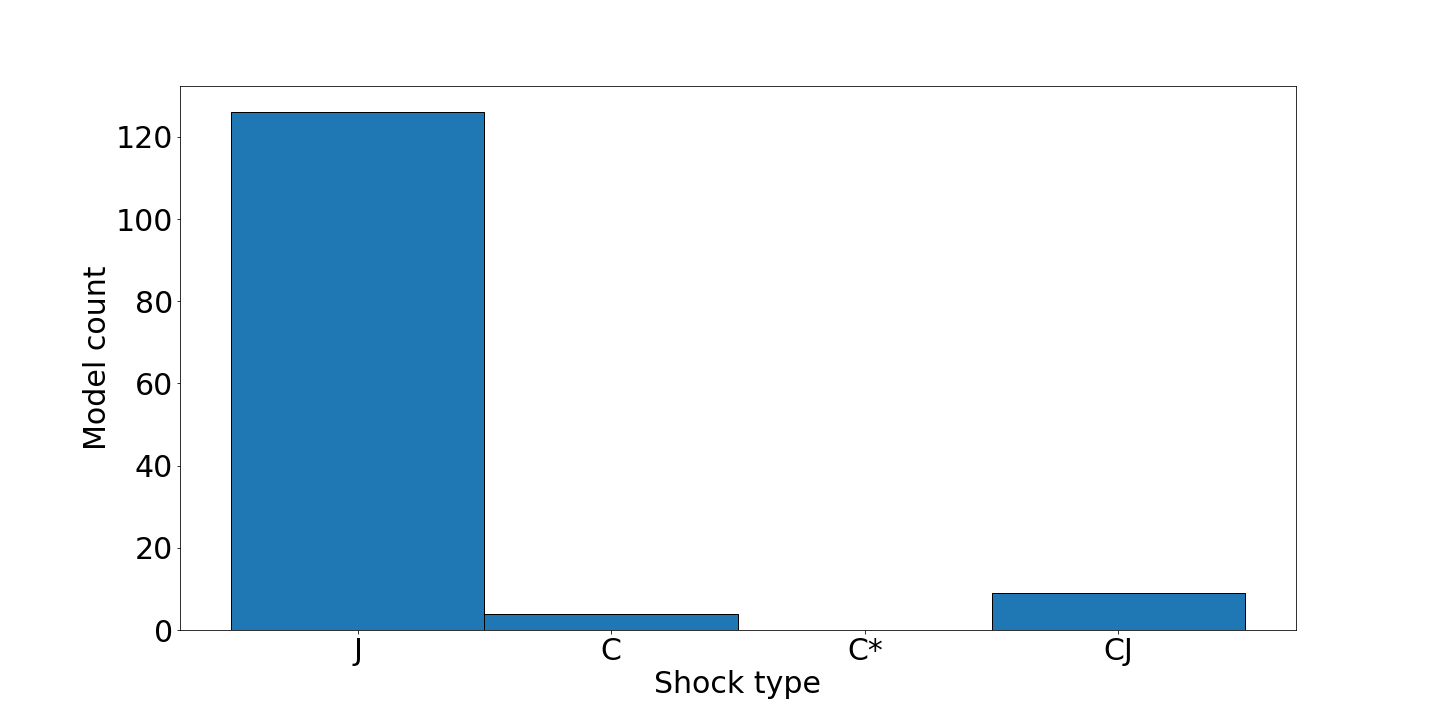}
    \caption{Histograms of models from \citet{kristensen2023} in agreement with the observed H$_2$ line ratios and upper limits. From top to bottom, the models show: pre-shock density, scaled transverse magnetic field, shock velocity, and shock type.}
    \label{fig: H2 models}
\end{figure}

\begin{figure}[h!]
    \centering
    \includegraphics[width=\columnwidth, trim={3cm 1cm 3cm 1cm, clip}]{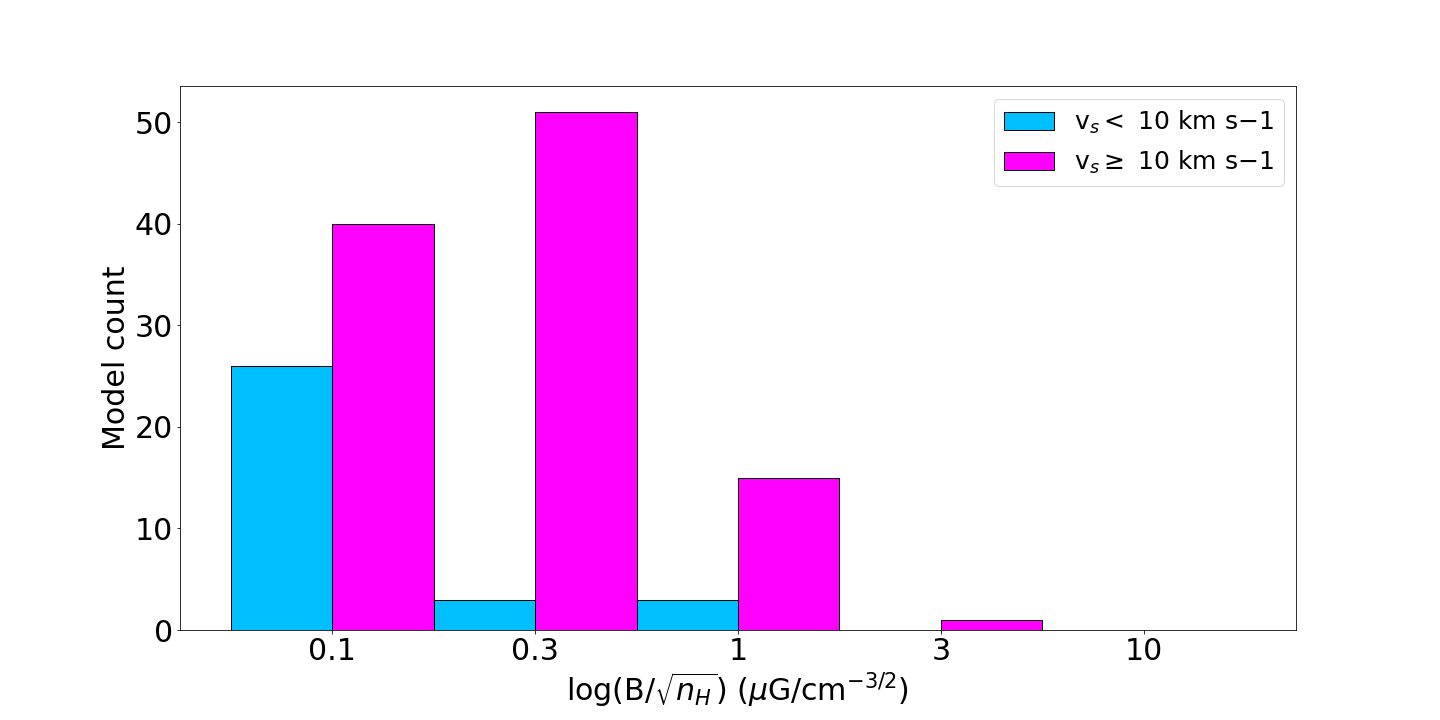}
    \caption{Histogram of models consistent with low ($\varv_s<$10 km s$^{-1}$, blue) and high ($\varv_s\geq$10 km s$^{-1}$, magenta) shock velocities relative to the input magnetic field strength.}
    \label{fig: H2 high+low vel}
\end{figure}
\end{appendix}
\end{document}